\documentclass[preprint, review, 12pt]{elsarticle}
\usepackage{multirow}
\usepackage{graphicx}
\usepackage{amsmath}
\usepackage{relsize}
\usepackage{longtable}
\usepackage{textcomp}
\usepackage{float}
\usepackage{amsmath}
\usepackage{textcomp}

\journal{Chemical Engineering Journal}

\biboptions{numbers,sort&compress}
\begin{document}

\begin{frontmatter}

	\title{CFD analysis of microfluidic droplet formation in non\textendash Newtonian liquid} 
	%%\tnotetext[mytitlenote]{Fully documented templates are available in the elsarticle package on \href{http://www.ctan.org/tex-archive/macros/latex/contrib/elsarticle}{CTAN}.}
	
	%% \author{Soma sekhara goud Sontti\fnref{myfootnote}}
	%% \address{Radarweg 29, Amsterdam}
	%% \fntext[myfootnote]{Since 1880.}
	
	%%  or include affiliations in footnotes:
	%% \author[mymainaddress,mysecondaryaddress]{Elsevier Inc}
	%% \ead[url]{www.elsevier.com}
	
	%%\author[mysecondaryaddress]{Arnab Atta\corref{mycorrespondingauthor}}
	\author{Somasekhara Goud Sontti}
	\author{Arnab Atta\corref{cor1}}
	\cortext[cor1]{Corresponding author. Tel.: +91 3222 283910}
	\ead{arnab@che.iitkgp.ernet.in}
	
	\address{Multiscale Computational Fluid Dynamics (mCFD) Laboratory, Department of Chemical Engineering, Indian Institute of Technology Kharagpur, West Bengal 721302, India}
	
	\begin{abstract}
		
		%The past few decades have seen the number of microfluidic investigations increasing due to numerous applications in  Lab\textendash on\textendash a\textendash Chip, chemical, biological. Most of these fundamental studies are focused on Newtonian fluids for their simplicity.
	A three-dimensional, volume-of-fluid (VOF) based CFD model is presented to investigate droplet formation in a microfluidic T-junction. Genesis of Newtonian droplets in non-Newtonian liquid is numerically studied and characterized in three different regimes, viz., squeezing, dripping and jetting. Various influencing factors such as, continuous and dispersed phase flow rates, interfacial tension, and non-Newtonian rheological parameters are analyzed to understand droplet formation mechanism. Droplet shape is reported by defining a deformation index. Near spherical droplets are realized in dripping and jetting regimes. However, plug shaped droplets are observed in squeezing regime. It is found that rheological parameters have significant effect on the droplet length, volume, and its formation regime. The formation frequency increases with increasing effective viscosity however, the droplet volume decreases. This work effectively provides the fundamental insights into microfluidic droplet formation characteristics in non-Newtonian liquids.  
	\end{abstract}
	
	
\end{frontmatter}

\section{Introduction}
In recent years, droplet\textendash based microfluidics offer a wide range of applications in the fields of lab\textendash on\textendash a\textendash chip, chemical, biological and nanomaterial synthesis \cite{liaq-2008,breisig-2014,knauer2011,xu2008,basova-2015,xua-2016}. In a microfluidic device, each droplet provides a compartment microreactor in which species transport or reactions can occur \cite{nisisako2004,plouffe2016,tsaoulidis2015,wans-2012,night-2013,zhangh-2014,niuda-2015,niu-2016}. Recently, there has been a rapid development on emulsion generation in microfluidic devices \cite{zhu2017}. Non-Newtonian liquid in multiphase system has also become a significant area of research in microfluidics due to its paramount importance in biomedical engineering. Droplets generated in non\textendash Newtonian medium are frequently encountered in biochemical and drug delivery applications. However, precise control over the droplet size is required to deliver accurate dosing of a drug or chemical reactant \cite{park-2015}. Therefore, monodisperse droplets are highly desirable in several areas of microfluidics \cite{li2016}. Various microfluidic devices, such as, T\textendash junction, flow-focusing, and co-flowing devices are available for generating droplets in microchannel by shearing dispersed phase with a continuous stream of liquid \cite{wang2014,christop-2007,barou-2010,gu-2011,huerre2014,zhu2017}. Microfluidic T\textendash junction finds wide spread utilization compared to other devices owing to its simplicity in geometric configuration and superior control over droplet size. In a T\textendash junction, cross\textendash flowing continuous and dispersed phase streams meet at the junction and the consequent shear force leads to droplet formation. The droplet size and frequency are guided by continuous and dispersed phase flow rate ratio, as well as by adjusting viscous and interfacial tension forces. T\textendash junction microchannel can be operated in three regimes namely, squeezing, dripping and jetting which are active functions of the primary driving forces acting on the system such as flow rate ratio, Capillary number (Ca), pressure gradient across the droplet, and wetting properties of the channel surface. \citet{thor-2001} initially reported generation of water droplets in oil using a T\textendash junction microchannel. \citet{nisi-2002} experimentally investigated the controlling parameters of droplet formation, size and frequency in an oil\textendash water system. Numerous researchers also proposed several strategies for droplet formation \citep{tice-2004,tan-2004,roach-2005, dendukuri-2005}. 

\citet{garstecki-2006} studied the droplet and bubble formation mechanism in a T-junction and proposed a power law correlation for predicting their sizes. Subsequently, several research works were carried out to develop scaling laws in forecasting the droplet length for different inlet configuration and fluid properties \citep{xu-2008,gupta-2009,guptaa-2010,wang-2011,fu-2012,fu2012bre, fu2016as}. With various surfactants and their concentrations, \citet{xu-2006} showed that adjusting interfacial tension and the wetting properties could lead to ordered or disordered two-phase flow patterns. \citet{bash-2011} numerically addressed similar issues and commented on active control of two\textendash phase flow pattern by altering interfacial tension and wetting properties. \citet{van-2006} computationally modeled droplet formation at the mesoscale using Lattice Boltzmann method (LBM). \citet{wang-2011} and \citet{riaud-2013} investigated the droplet formation mechanism by LBM with simple modifications in a T\textendash junction and shearing plates. \citet{yang-2013} also applied LBM to analyze the droplet formation and cell encapsulation process where three flow regimes were illustrated and the droplet shapes are reported in each regime, such as, plug shape in squeezing, and bullet shape in dripping as well as in jetting regimes. \citet{Wangg-2011} discussed the generation of monodisperse droplets using capillary embedded T\textendash junction device and described its dependence on \textit{Ca}, viscosity ratio, and dispersed phase flow rate. \citet{de-2008} computationally illustrated different flow regimes using phase-field method. \citet{chris-2008} and \citet{fu2010squ} analyzed bubble formation in microfluidic T-junctions and its transition from squeezing to dripping mechanism. \citet{fu2010squ} also proposed scaling laws in terms of flow rate ratio and \textit{Ca} for the estimation of bubble sizes in various regimes. Additional insights to the dispersed and continuous phase pressure profiles in droplet breakup process can be gained from the research of \citet{sivasamy-2011}. Effects of liquid viscosity and interfacial tension on droplet formation were investigated by \citet{chen-2016} which showed that the period of slug formation increases with increasing interfacial tension. \citet{raj-2010} studied droplet formation in T-junction and Y\textendash junction microchannels using volume-of-fluid (VOF) method and analyzed the effect of flow rate ratio, liquid viscosity, interfacial tension, channel size, and wall adhesion properties on slug length for Newtonian liquids. \citet{fu-2012} investigated oil droplet formation in a flow\textendash focusing microchannel and proposed scaling laws to predict the droplet length. Benefiting from the control over shape and size, few studies are also reported in T\textendash junction device to produce spherical and plug shaped nanoparticles \citep{zhang-2014,xua-2016,li2013}.   

%The results revels the pressure evaluation in both continuous and dispersed phase with time. It's clearly stated that, disperses phase pressure increases with change in continuous phase pressure difference. 

%Recently \citet{}  for the first time studied the formation of ionic liquid droplets in T\textendash junction to the synthesis of gold and silver nanoparticles.

Interestingly, most of the reported research are concerned with the droplet formation mechanism and flow regimes in Newtonian fluids, while in several applications, liquid phases are likely to exhibit complex behaviors, such as non\textendash Newtonian properties. For example, it is apparent that in most cases of oil-water emulsions, oil phase is usually considered as the Newtonian liquid during modeling of such flow systems. However, studies on droplet formation in non\textendash Newtonian media is scarce. \citet{abatea-2011} studied monodisperse microparticle formation in non\textendash Newtonian polymer solutions in a flow\textendash focusing device. \citet{arratia2008p} reported polymeric filament thinning and breakup of Newtonian and viscoelastic liquids in a flow\textendash focusing microchannel. The results showed different breakup mechanisms for Newtonian and polymeric liquids having same viscosity. This phenomena was attributed to the rheological difference between the two types of liquids. \citet{qiu-2010} numerically investigated the droplet formation in non\textendash Newtonian liquids in a cross\textendash flow microchannel. It is apparent from their findings that rheological parameters of non-Newtonian fluid significantly influence the formation mechanism and size of droplets. \citet{aytoun-2013} experimentally examined the droplet pinch\textendash off dynamics in Newtonian, yield stress, and shear thinning fluids. \citet{fufu-2015} studied the flow patterns at the mesoscale in Newtonian and non\textendash Newtonian fluids using T\textendash junction millichannel. However, in-depth understanding of droplet formation and flow regimes in non\textendash Newtonian fluids is still lacking at the microscale. Therefore, analyses considering the liquid phase as non\textendash Newtonian fluid is imperative and desirable. In this study, we develop a CFD based model to understand the droplet formation in a T\textendash junction microchannel considering the continuous phase as non\textendash Newtonian liquid. We describe the role of rheological properties, interfacial tension, and flow rate ratio on droplet characteristics, in terms of formation mechanism, droplet length, volume, velocity, and its shape. These understandings can significantly benefit in setting guidelines to control droplet size and shape in non\textendash Newtonian liquids. 

%
%Few researchers  studied droplet formation in non\textendash Newtonian fluids. To understand the rheological properties of non\textendash Newtonian fluids, extensive experiments are required. Therefore, it is imperative to develop the numerical
%simulation method which provides droplet formation mechanism in non\textendash Newtonian fluids. It is hard to obtain all details of underlying physics in non\textendash Newtonian investigations in a microfluidic system. Because of these reasons, a well verified model can serve as an alternative to rigorous experimentation. 

% This observation reveals different flow regimes (dripping, squeezing and jetting) which are essential, where interfacial tension and shear stress have a dominating effect on droplet length for power\textendash law fluids.  These  numerical results would give new insights droplet formation in   Power\textendash law fluids using T\textendash junction microchannels.  

\section{Problem formulation} 

\begin{figure}[H]
	\centering
	\includegraphics[width=0.80\textwidth]{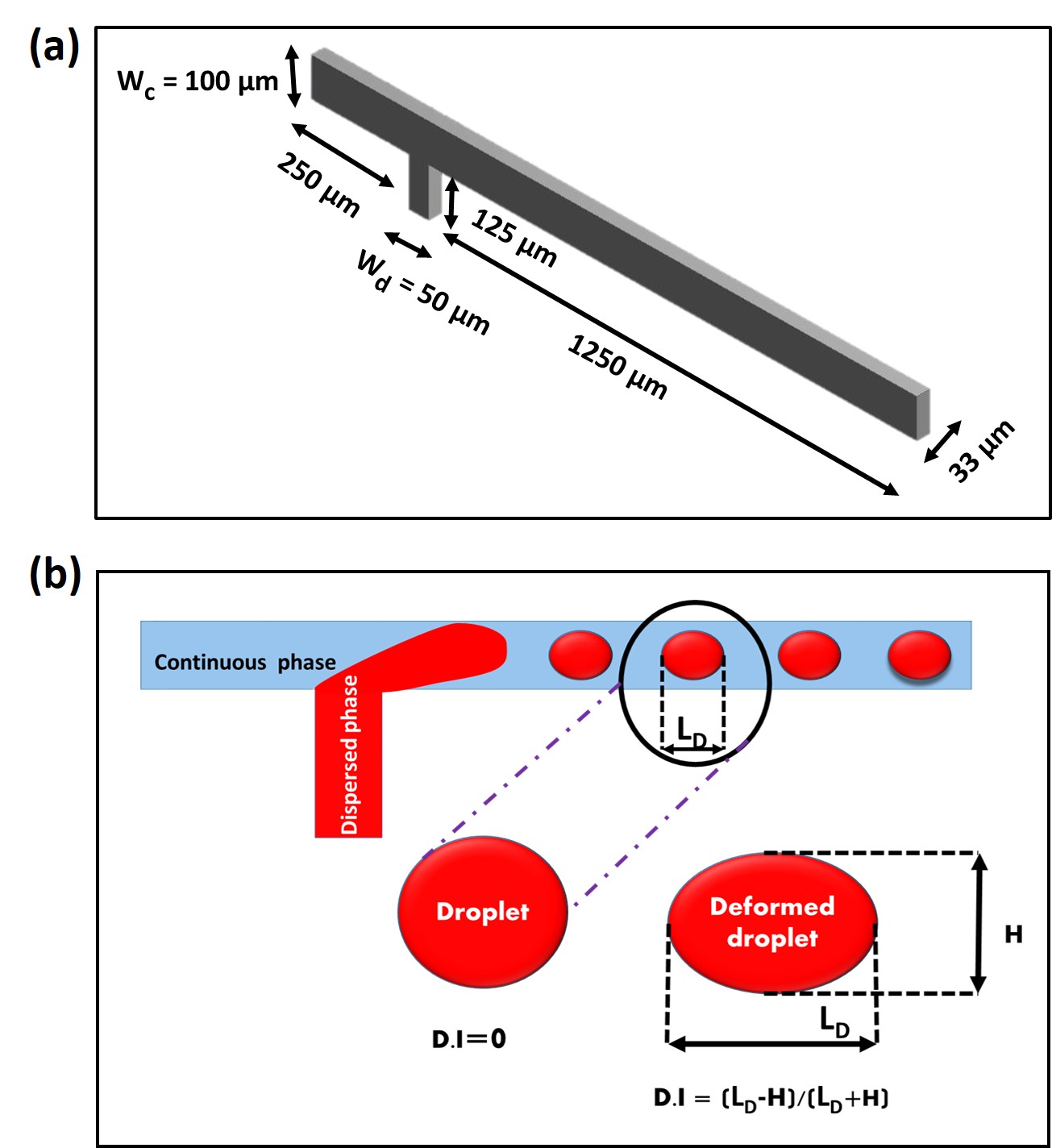}
	\caption{\label{fig:M}(a) Considered 3D T\textendash junction geometry, and (b) 2D schematic of droplet formation in a T\textendash junction microchannel.}
\end{figure}

A three\textendash dimensional  T\textendash junction microchannel, as shown in Fig. \ref{fig:M}a, is considered to investigate the droplet formation in a non-Newtonian liquid which flows through the main channel having a cross-section of 100~$\mu m$ $\times$ 33~$\mu m$. Newtonian dispersed phase is introduced through a perpendicular channel of 50~$\mu m$ $\times$ 33~$\mu m$. At the merging junction of these two liquids, continuous phase shear force acts on the dispersed phase and the consequent pressure gradient results in formation of droplets that flow through downstream of the channel (Fig. \ref{fig:M}b). Typically, the droplet formation characteristics are governed by the complex interactions between the two phases resulting from forces like viscous, interfacial, shear and pressure gradient in the channel. These forces are substantially influenced by the liquid properties and flow rates. To quantify droplet shape, a deformation index ($D.I$) is calculated as shown in Fig. \ref{fig:M}b, which indicates an undeformed state at $D.I=0$ whereas, $D.I>0$ suggests deformed droplets with length greater than its height.  
%A length of 1250  $\mu m$ from the merging junction to the channel outlet is chosen in this study and model geometry with dimensions as shown in Fig.\ref{fig:M} (a).

%\noindent Typically, the droplet formation in T\textendash junction microchannel, the continuous phase ($Q_o$) fluid flows through the main channel and dispersed phase ($Q_w$) fluid is introduced perpendicular to the continuous phase fluid, to generates droplets in T\textendash junction. 

%Droplet formation time changes sequentially due to the impact of shear force on the dispersed phase. In this work, the shape of the droplet is represented with Droplet deformation index (D.I). When D.I approaches to zero, the shape of the droplet is spherical (undeformed),  Schematic depiction of droplet formation and significance of D.I  as illustrated in  Fig.\ref{fig:M}b.

\subsection{Governing equations}
\subsubsection{Equations of continuity and momentum}
In this work, incompressible two\textendash phase (oil\textendash water) flow is considered, where oil and water are continuous, and dispersed phases, respectively. The interface is tracked by VOF method, which solves a single set of conservation equations for both phases, as follows \cite{Ranadebook}:  

\textbf{Equation of continuity:}
\begin{equation}
\label{eq:con_eqn}
\nabla. (\rho  \vec{ U}) = 0
\end{equation}

\textbf{Equation of motion:}

\begin{equation}
\label{eq:mom_eqn}
\frac{ \partial (\rho \vec{ U })}{ \partial t} + \nabla.( \rho \vec{ U } \vec{ U }) = - \nabla P + \nabla.\overline{\overline \tau} + \vec{ F}_{SF}
\end{equation}

where $\vec{ U }$, $P$, $\rho$, and $\overline{\overline \tau}$ are velocity vector, pressure, volume averaged density, and stress tensor, respectively. For incompressible Newtonian fluids, the shear stress is proportional to the rate\textendash of\textendash strain tensor ($\dot{{\gamma}}$), described by:

%\begin{empheq}[box=\tcbhighmath]{align}
%\centering
%\label{eq:tau}
%%\tau = k\left(\frac{\partial v}{\partial y}\right)^{n-1}\left(\frac{\partial v}{\partial y}\right)=\mu_{0}\frac{\partial v}{\partial y}
%% = \eta (\nabla \vec {U} + \nabla { \vec {U} } ^{T})  
%\overline{\overline \tau} = \eta \overline{\overline D}
%\end{empheq}

\begin{equation}
\overline{\overline \tau}= \eta \dot{ \gamma } = \eta (\nabla \vec {U} + \nabla { \vec {U} } ^{T})
\end{equation}

%
% \hl{Where $\overline{\overline D}$ is defined by}
%\begin{empheq}[box=\tcbhighmath]{align}
%\centering
%\label{eq:defromation}
%\overline{\overline D}= \frac{\partial U_{j}}{\partial X_{i}} +\frac{\partial U_{i}}{\partial X_{j}}
%\end{empheq}

where $\eta$ is the volume-averaged viscosity. The volume-averaged properties are defined in terms of oil ($\alpha_o$) and water ($\alpha_w$) volume fractions, as follows:

\begin{equation}
\label{eq:rho_eqn}
\rho  =\alpha _{o} \rho _{o} + (1-\alpha _{w}) \rho _{w}  
\end{equation}
\vspace{-1cm}
\begin{equation}
\label{eq:vis_eqn}
\eta =\alpha _{o} \eta_{o} + (1-\alpha _w) \eta_{w}
\end{equation}

\textbf{Equation of volume fraction:}

The volume fraction of each liquid phase is calculated by solving the following equation:   

\begin{equation}
\label{eq:vol_eqn}
\frac{\partial \alpha_q}{\partial t} + \vec{U}. \nabla \alpha_q = 0
\end{equation}

where the subscript $q$ refers to either oil ($o$) or water ($w$) phase. In each computational cell, the volume fractions of all phases are conserved by $\sum  \alpha _{q} =1$. For $\alpha_{q}=0$, the reference cell is assumed to be devoid of the $q$th phase, and $\alpha_{q}=1$ indicates that the cell is  completely filled with $q$th phase. Consequently, the interface between two phases is identified by marking the cell with volume fraction range $0 < \alpha_q < 1$.

\subsubsection{Surface tension force}

\noindent The continuum surface force (CSF) model \citep{brack-1992} is used to define the volumetric surface tension force ($F_{SF})$ term in Eq. \ref{eq:mom_eqn}, as follows:

%\begin{empheq}[box=\tcbhighmath]{align}
%\label{eq:source_eqn}
%\vec{F} _{SF} = \sigma  \begin{bmatrix} (\frac{\rho  \kappa_{N}  \nabla \alpha_o}{ \frac{1}{2 } ( \rho_{o}+  \rho_{w})})  \end{bmatrix}    
%\end{empheq}

\begin{equation}
\label{eq:source_eqn}
\vec{F} _{SF} = \sigma  \begin{bmatrix} \frac{\mathlarger{\rho } \mathlarger{\kappa_{N}} \mathlarger{\nabla \alpha_o}}{ \mathlarger{\frac{1}{2 }} (\mathlarger{\rho_{o}+  \rho_{w}) }}  \end{bmatrix}    
\end{equation}

\noindent where $\kappa_{N}$ is the radius of curvature and $\sigma$ is the coefficient of surface tension. The  interface curvature ($\kappa_{N}$) is calculated in terms of unit normal $\hat{N}$, as:

\begin{equation}
\label{eq:normal_eqn}
\kappa_{N} = - \nabla  .  \hat{N}= \frac{1}{|\vec{ N}|}  \begin{bmatrix} \big( \frac{\vec{ N}}{ |\vec{ N}|} . \nabla\big) |\vec{ N}| - \big( \nabla  . \vec{ N}\big) \end{bmatrix}, \text {where} ~\hat{N} = \frac{\vec{ N}}{ | \vec{ N} | }
\end{equation}

%\begin{centering}  
%	
%	$\hat{N} = \frac{\vec{ N}}{ | \vec{ N} | } $\\
%	
%\end{centering}

\noindent In VOF formulation, surface normal, $N$, is expressed as the gradient of phase volume fraction at the interface which can be written as: 

\begin{equation}
\centering
	\vec{ N}= \nabla  \alpha _{q} 	
\end{equation}

This surface tension force is implemented by the piecewise-linear interface calculation (PLIC) scheme that provides accurate calculation of curvatures for reconstruction of the interface front \cite{fluent,guo2015}. Wall adhesion effect is also taken into consideration by defining a three-phase contact angle at the channel wall ($\theta_{W}$). Accordingly, the surface normal at the reference cell next to the wall is given by:

\begin{equation}
\label{eq:normal1_eqn}
\hat{N}=  \hat{N}_{W}  cos \theta_{W}  +  \hat{M}_{W} sin \theta _{W} 
\end{equation}

\noindent where $\hat{N}_W$ and $\hat{M}_W$ are the unit vectors normal and tangential to the wall, respectively \cite{fluent}. In this work, a static contact angle condition is specified which is assumed to be independent of the moving contact line and the velocity \cite{raj-2010}. The surface normal one cell away from the wall along with the contact angle govern the local curvature of the surface, which is utilized in the calculation of the surface tension force (Eq. \ref{eq:source_eqn}) to determine the body force term in Eq. \ref{eq:mom_eqn}. For real surfaces, the contact angle varies dynamically between an advancing and a receding contact angle. If the contact angle remains within the range of advancing and receding angles, the contact line does not move \cite{Choi2008}. Typically, the use of static contact angle has been proved to be adequate for analyzing the flow behavior in microchannels \cite{van-2006,wang-2011,hoang2013,zhang2014loc,kagawa2014}. Nonetheless, the cases of dynamic contact angles can be resolved by defining a level set function \cite{sussman1994} and coupled with the VOF model. However, such investigation is beyond the scope of this present work and can be found elsewhere \cite{Choi2008,lee2008}. 

%The contact angle, $\theta _{ \omega }$ is the angle between the wall and the tangent to the interface.

\subsubsection{Constitutive equation of continuous phase}

For non\textendash Newtonian liquids, the shear stress can be written in terms of a non\textendash Newtonian viscosity:

%\begin{center}
%	$\dot{ \gamma } = \sqrt{2 \big( \vec{D} :  \vec{D}  \big) } $
%\end{center}    

%Where $\vec{D}$ is strain rate tensor, can be as follows

%\begin{center}
%	$\vec{D}= \frac{1}{2} \big( \bigtriangledown \vec{ \nu } + \bigtriangledown   \vec{ \nu } ^{T}  \big) $ 
%\end{center} 
%
%Therefore, non\textendash Newtonian fluid  apparent viscosity can be obtained  from above equations.

\begin{equation}
\centering
\label{eq:nnvis}
\overline{\overline \tau} = \eta(\dot{\gamma})\dot{\gamma}
\end{equation}

where  $\eta$ is a function of all three invariants of the rate\textendash of\textendash deformation tensor. However, in power\textendash law model, the non\textendash Newtonian liquid viscosity ($\eta$) is considered to be a function of only shear rate ($\dot{ \gamma }$).
	\begin{equation}
	\label{eq:nnvis_eqn}
	\eta(\dot{\gamma}) =K \dot{\gamma }^{n-1} 
	\end{equation} 
	
where $K$ and $n$ are the consistency and power-law indices, respectively. The local shear rate ($\dot{ \gamma}$) is related to the second invariant of $\overline{\overline D}$ and is expressed as \cite{fluent}:

\begin{equation}
\label{eq:shearrate}
\dot{ \gamma } = \sqrt{ \frac{1}{2} (\nabla \vec {U} + \nabla { \vec {U} } ^{T})_{ij} (\nabla \vec {U} + \nabla { \vec {U} } ^{T})_{ji}} 
%\dot{ \gamma } = (\nabla \vec {U} + \nabla { \vec {U} } ^{T}) 
\end{equation} 

%To understand the Newtonian droplet formation phenomena in non\textendash Newtonian liquid phase, power-law model is considered for calculating the apparent viscosity ($\eta$) that is expressed as:
%\begin{equation}
%\label{eq:nnvis_eqn}
%\eta =K \dot{\gamma }^{n-1} 
%\end{equation} 

%where $\dot{ \gamma}$, $K$ and $n$ are the  local shear rate, consistency coefficient and flow index, respectively \cite{fluent}. 

\subsection{Implementation in numerical model}

Aforementioned time dependent governing equations are solved in a CFD solver (Ansys Fluent 17.0) based on finite volume method. Pressure implicit with splitting of operators (PISO) algorithm \cite{issa1986} is used to resolve the pressure\textendash velocity coupling in momentum equation. The spatial derivatives are discretized using quadratic upstream interpolation for
convective kinetics (QUICK) scheme \cite{leonard1979}. To avoid spurious currents as a result of mismatch between pressure and surface tension force discretization, the pressure staggering option (PRESTO) is employed for pressure interpolation \cite{fluent, guo2015}. The geometric reconstruction scheme is adopted to solve the volume fraction equation. First-order implicit method is applied for the discretization of temporal derivatives. Subsequently, variable time step and fixed Courant number ($Co=0.25$) are considered for simulating the governing equations. 

Constant velocity for both continuous and dispersed phases are imposed at the inlets and outflow boundary condition is specified at the outlet. It is assumed that continuous phase completely wets the channel wall and the solid walls are set to no\textendash slip boundary condition with a static contact angle of $135^{0}$. Structured hexahedral meshes with Green\textendash Gauss \textit{node} based schemes \citep{seifollahi2008} are used in the computational domain to accurately calculate the gradients and to overcome calculation inaccuracies resulting from spurious currents, as recommended by \citet{gupta2009}. Moreover, the effect of mesh element size is initially investigated for analyzing numerical diffusion that mainly stems from poor meshing scheme. Thereafter, the model is validated for the droplet length estimation with the reported experimental data of \citet{garstecki-2006}. It has been realized that for more than 2.5 times of mesh elements (comparing between 137,176 and 360,640 elements), the difference in droplet length is nearly 1\%. Consequently, for the considered geometry all simulations are performed with element size of 4 $\mu m$ (137,176 elements) to optimize the computational time. 

%The properties of oil and water used in model validation are provided in Table \ref{T1}. 
%  
%\begin{table}[!ht]
%	\centering
%	 \footnotesize
%\caption{Physical properties of liquid\textendash liquid systems used in model validation}
%\vspace{10px}
%\label{T1}
%\begin{tabular}{clllc}
%		\hline
%		\multicolumn{1}{l}{Refrence}            & Fluid & \begin{tabular}[c]{@{}l@{}}Density \\ ($kg/m^3$)\end{tabular} & \begin{tabular}[c]{@{}l@{}}Viscosity \\ (Pa.s)\end{tabular} & \multicolumn{1}{l}{\begin{tabular}[c]{@{}l@{}}Interfacial tension\\  (N/m)\end{tabular}} \\ \hline
%		\multirow{2}{*}{\citet{garstecki-2006}} & Oil   & 930                                                           & 0.01                                                        & \multirow{2}{*}{0.0365}                                                                   \\ \cline{2-4}
%		& Water & 1000                                                          & 0.001                                                       &                                                                                           \\ \hline
%	\end{tabular}
%\end{table}

\subsection{Model validation}

\begin{figure}[!ht]
	\centering
	\includegraphics[width=1\textwidth]{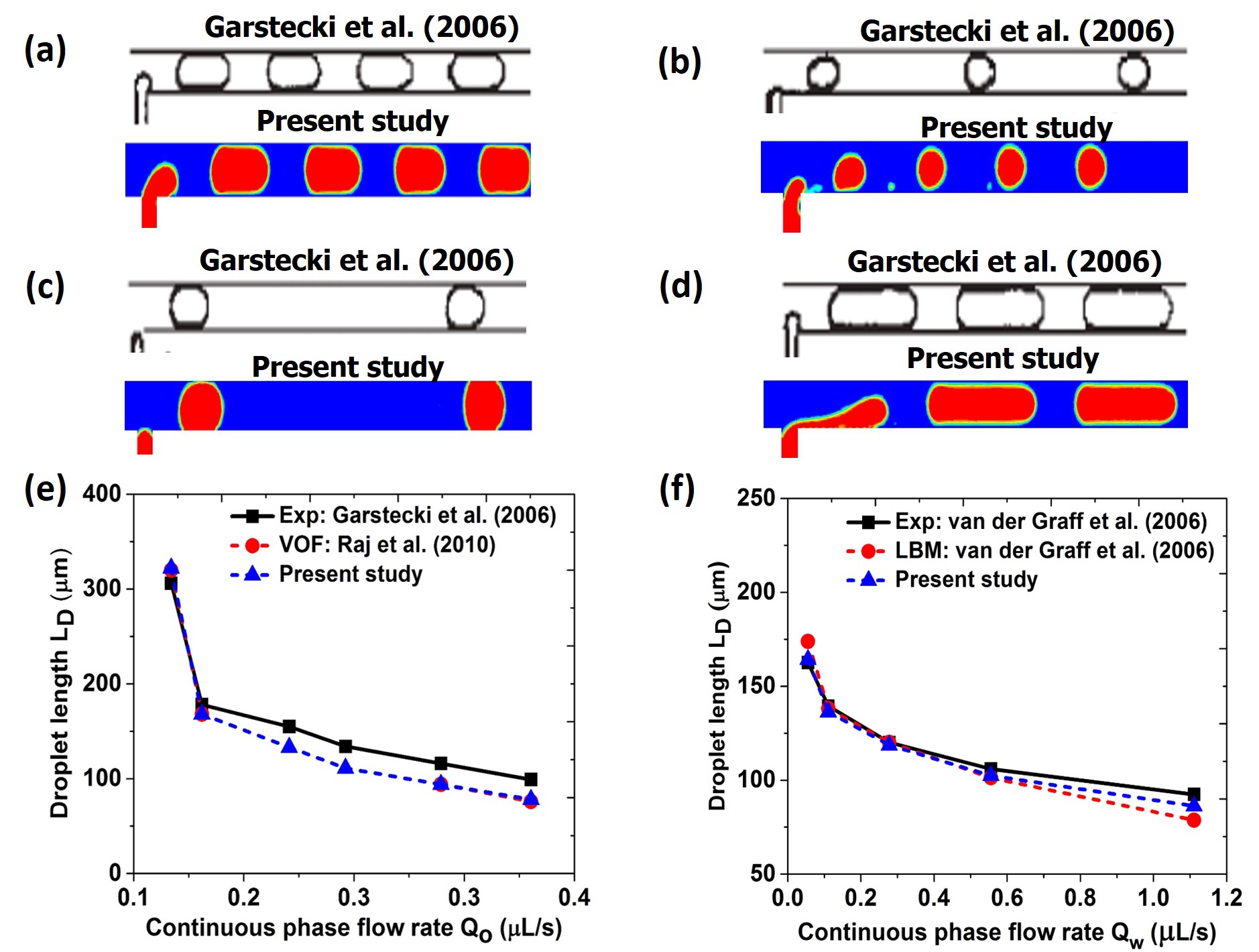}
	\caption{\label{fig:M1}Comparison of the model predictions with the experimental results of \citet{garstecki-2006} (figures reprinted  with permission from the publisher, Royal Society of Chemistry) for (a) $Q_w = 0.14$ $\mu L/s$ and $Q_o$ = 0.124 $\mu L/s$, (b) $Q_w = 0.14$ $\mu L/s$ and $Q_o$ = 0.408 $\mu L/s$, (c) $Q_w$ = 0.004 $\mu L/s$ and $Q_o$ = 0.028 $\mu L/s$, (d) $Q_w$ = 0.006 $\mu L/s$ and $Q_o$ = 0.028 $\mu L/s$. Comparison of droplet length with (e) experimental results of \citet{garstecki-2006} and numerical predictions by \citet{raj-2010} at $Q_w$ = 0.14 $\mu L/s$, $\sigma$ = 0.0365 N/m, $\eta_o = 0.01~ Pa.s$, $\eta_w = 0.001~ Pa.s$, and (f) experimental as well as LBM results of \citet{van-2006} at $Q_w$ = 0.055 $\mu L/s$, $\sigma$ = 5 mN/m, $\eta_o = 6.71~ mPa.s$ and $\eta_w= 1.95~ mPa.s$.}
\end{figure}

Fig. \ref{fig:M1} shows the comparison of model predictions with the experimental observations of \citet{garstecki-2006}. At a constant water flow rate ($Q_w = 0.14$ $\mu L/s$), the droplet size variation with increasing oil flow rate ($Q_o = 0.124-0.408$ $\mu L/s$) is illustrated in Fig. \ref{fig:M1}a and Fig. \ref{fig:M1}b. Similarly, Fig. \ref{fig:M1}c and Fig. \ref{fig:M1}d describe the effect of increasing dispersed phase (water) flow rate ($Q_w = 0.004-0.006$ $\mu L/s$) for $Q_o =0.028~\mu L/s$. Fig. \ref{fig:M1}e portrays the quantitative comparison of droplet length estimation with the experimental observation by \citet{garstecki-2006} which shows the maximum error of 18\% at the highest continuous phase flow rate. However, the model results are found to be identical with the numerical prediction by \citet{raj-2010}. The developed model is further verified with the experimental results of \citet{van-2006} for asserting its accuracy in droplet length prediction. In this case, oil droplet formation is simulated by changing the continuous phase (water) flow rate in a T\textendash junction microchannel and a maximum deviation of 6\% from the experimental results is observed, as shown in Fig.~\ref{fig:M1}f. This validation also establishes the efficacy of the developed model to forecast the droplet length better than the LBM simulations, reported by \citet{van-2006}.

\section{Results and discussion}

Armed with fairly validated model, the droplet formation mechanism and its behavior in non-Newtonian liquids are systematically investigated for different operating conditions by varying the continuous and dispersed phase velocities ($Q_o$ and $Q_w$, respectively). The influence of various rheological parameters, namely, power\textendash law index ($n$), consistency index ($K$), and, interfacial tension ($\sigma$) are elaborated in the following sections.   
%
%In this study, we alter the continuous phase (Oil) as a non\textendash Newtonian fluid while keeping other properties constant and examine the effect of non \textendash Newtonian fluid on droplet formation and dimensionless droplet length. The primary focus of this work is to reveal the mechanism of droplet formation in non\textendash Newtonian fluids.

\subsection{Effect of power\textendash law index}

\begin{figure}[!ht]
	\centering
	\includegraphics[width=1\textwidth]{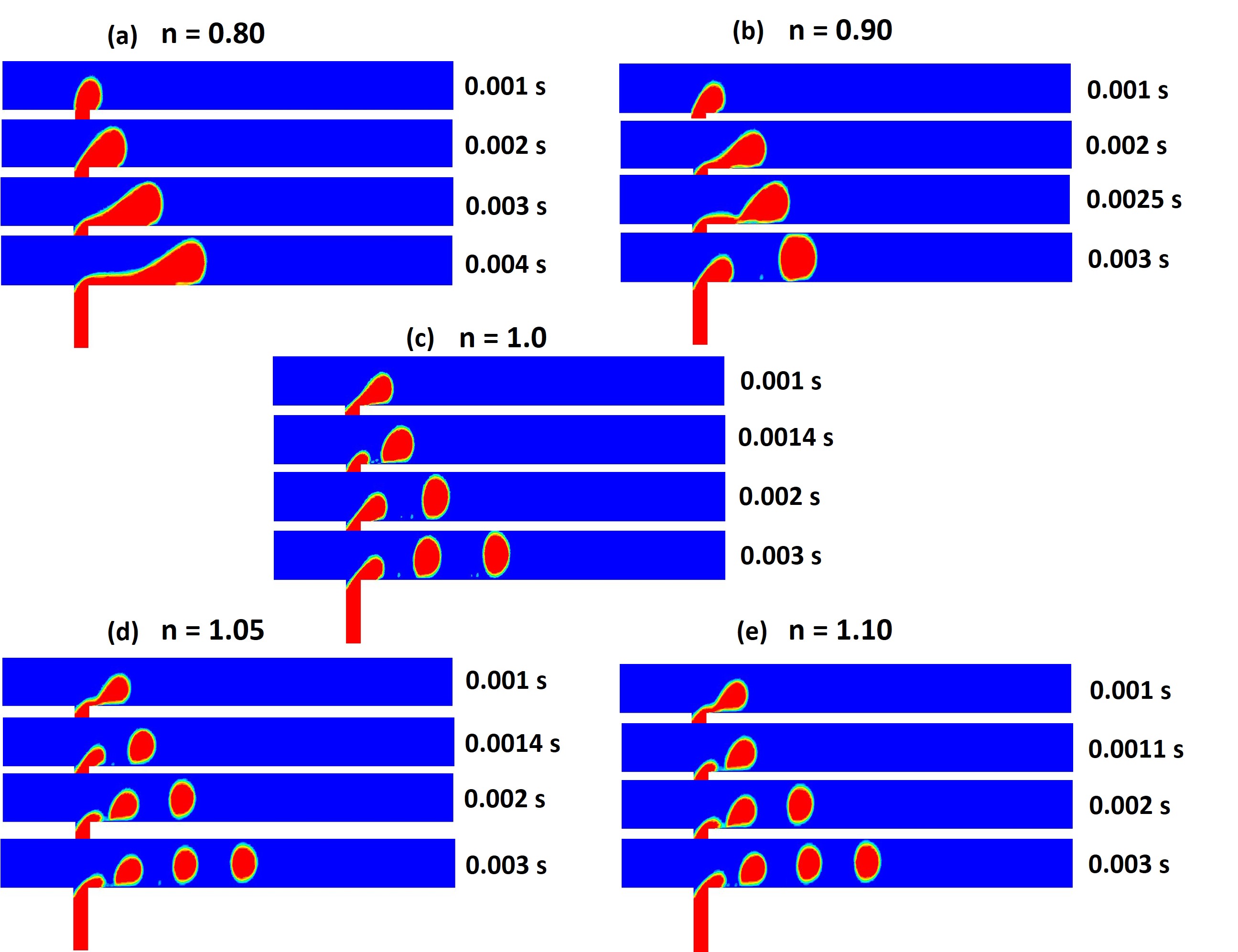}
	\caption{\label{fig:n1}Droplet formation mechanism for Newtonian and power-law liquids at a fixed operating condition of $Q_o = 0.408 $ $\mu L/s$, $Q_w = 0.14 $ $\mu L/s$, K= 0.01 $Pa.{s^n}$, $\eta_w = 0.001$ Pa.s and $\sigma$ = 0.0365 N/m.}
\end{figure}

Fig. \ref{fig:n1} portrays the temporal evolution of droplet formation phenomena in different power\textendash law liquids, obtained by varying $n$ from 0.80 to 1.10.
It can be understood from Fig. \ref{fig:n1} that at a fixed operating condition, droplet formation time decreases with increasing power\textendash law index. This is attributed to the interplay between viscous and interfacial forces. Although the droplet formation process is similar for $n=1.0$ and $n=1.05$ (mainly due to the small change in power-law index) till 0.0014 s, it shows a significant change at 0.003 s where three droplets are formed for $n=1.05$. With increasing $n$, the effective viscosity ($\eta_{eff}$) of the continuous phase (oil) is enhanced, which in turn, imparts higher viscous resistance and helps in rapid detachment of droplets. In the cases of shear thinning liquids ($n<1$), interfacial force plays a dominant role and the detachment of dispersed phase at the merging junction is delayed. 

\begin{figure}
	\centering
	\includegraphics[width=1\textwidth]{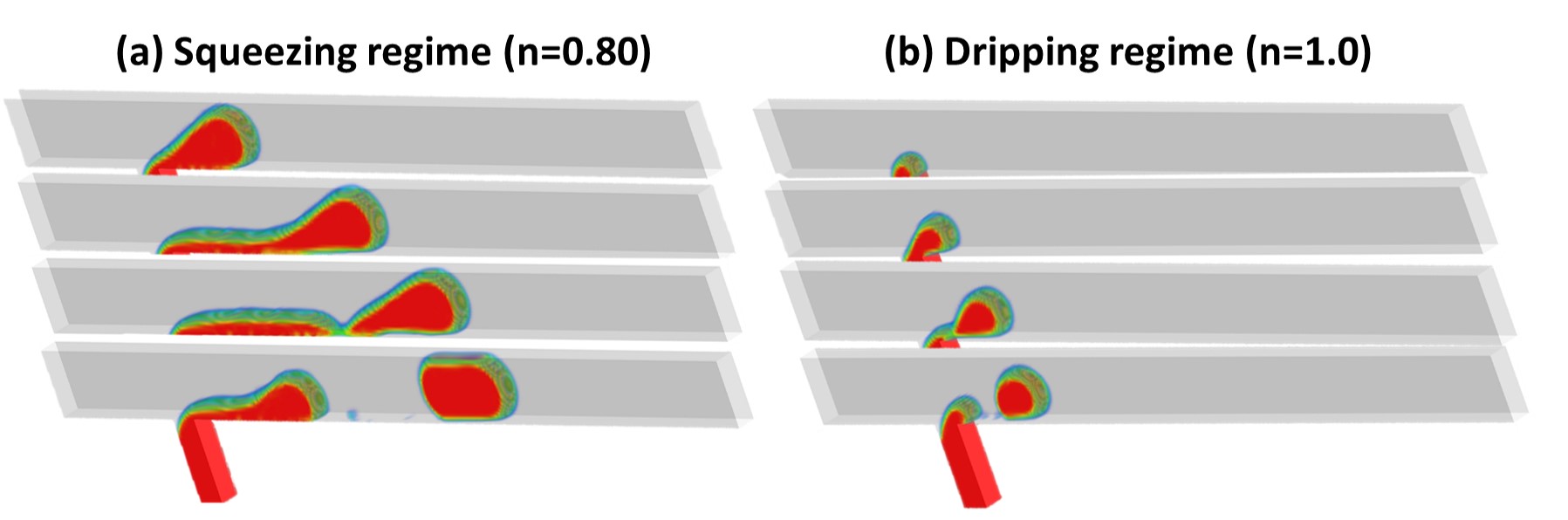}
	\caption{\label{fig:n3} Droplet formation mechanism (a) squeezing regime ($n$=0.80) and (b) dripping regime ($n$=1.10)  at a fixed operating condition of $Q_o = 0.408 $ $\mu L/s$, $Q_w = 0.14 $ $\mu L/s$, K= 0.01 $Pa.{s^n}$, $\eta_w = 0.001$ Pa.s and $\sigma$ = 0.0365 N/m.}
\end{figure}

Moreover, for shear thinning liquid, squeezing regime is observed, where the dispersed phase grows slowly and covers the entire flow area (Fig. \ref{fig:n3}a). This results in restricted flow of the continuous phase and consequently, pressure gradient in continuous phase across the droplet increases. Eventually, the formation of plug shaped droplets are formed once the gradient is sufficiently large to overcome the pressure inside the dispersed phase droplet. Dripping regime of droplet formation is observed for Newtonian and shear thickening liquids ($n\ge1$), where the droplet pinch off occurs at the merging junction, as depicted in Fig. \ref{fig:n3}b. In the dripping regime, viscous force that acts on the interface to snap off dispersed phase, typically dominates over the interfacial tension force \cite{zhu2017}. As a consequence of larger viscous and shear forces in Newtonian and shear thickening liquids with increasing $n$, Fig. \ref{fig:n1} shows rapid formation of droplets before its enlargement up to the channel top wall. It can be further noted that as the flow regime shifts from squeezing to dripping, droplet shape also changes from plug to spherical. 
%At fixed flow time, with increase in power\textendash law index number of droplets increase in the main channel for Newtonian and non \textendash Newtonian fluids as shown in Fig.\ref{fig:n1}.  Droplet pinched off at the neck for $n\textgreater 0.90$ but in the case of n=0.80 droplets are not formed until 3 $ms$ flow time as shown in Fig.\ref{fig:n1}. 
%\noindent Droplet length is characterized by dimensionless length ($L_D/W_c$) as the ratio droplet length ($L_D$) to continuous phase channel width ($W_c$) and droplet volume obtained by volumetric integral of the disperse phase. When dimensionless droplet ($L_D/W_c\textless1$) is less than unity, it indicates that the length of droplet is smaller than $W_c$. Similarly, $L_D/W_c\textgreater1$ denotes higher droplet length than  $W_c$. 

\begin{figure} [!ht]
	\centering
	\includegraphics[width=1\textwidth]{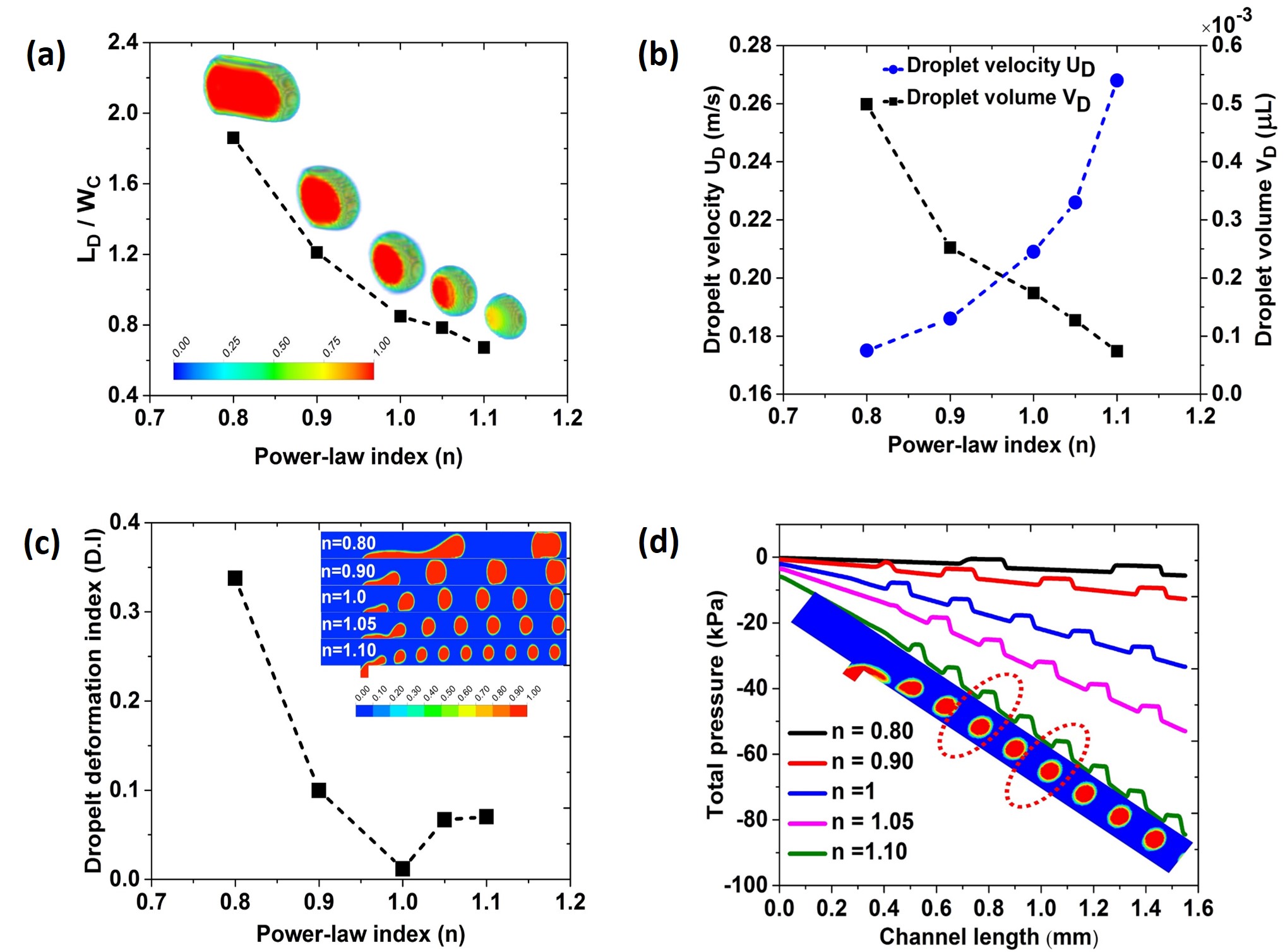}
	\caption{\label{fig:n2} Effect of power\textendash law index on (a) non-dimensional droplet length (b) droplet velocity and volume, (c)  deformation index ($D.I$), and (d) pressure profiles along the channel centerline at fixed $K= 0.01~Pa .s^n$, $\eta_w = 0.001~Pa.s$, $\sigma = 0.0365~N/m$, $Q_o = 0.408 ~\mu L/s$, and $Q_w = 0.14~\mu L/s$. }
\end{figure}
Fig. \ref{fig:n2}a shows that non-dimensional droplet length ($L_D/W_c$) decreases with increasing power \textendash law index ($n$), due to  increase in effective viscosity of continuous phase liquid. It is also apparent that in shear thinning liquids, elongated droplets are formed ($(L_D / W_c > 1)$
), whereas smaller droplets are formed for ($L_D / W_c < 1$) Newtonian and shear thickening liquids. Accordingly, droplet volume also decreases with increasing $n$, as shown in Fig. \ref{fig:n2}b. However, the droplet velocity is found to increase with increasing $n$. This can be ascribed to the change in film thickness and  flow profile that occur in continuous phase from shear thinning to thickening nature. Typically, the velocity profile is sharper for shear thickening liquids compared to a plug flow profile for shear thinning cases \citep{yoshi-2007}. Based on length and height, droplet shape is also quantified in terms of droplet deformation index ($D.I$), as illustrated in Fig. \ref{fig:n2}c. Plug shaped droplets are identified  for $n$\textless1 and almost spherical shaped droplets are observed for $n$=1 when all the other parameters such as $K$, interfacial tension, and flow rates were kept constant. For $n$\textgreater1, small deformation is realized from spherical shape due to higher viscous force. The pressure evolution along the channel length for different power\textendash law liquids is described in Fig. \ref{fig:n2}d. It can be seen from Fig. \ref{fig:n2}d that pressure drop increases with increasing power\textendash law index and each peak represents the droplet position (encircled by dotted red line) where there is a pressure difference before and after droplet detachment.
\begin{figure} [!ht]
	\centering
	\includegraphics[width=1\textwidth]{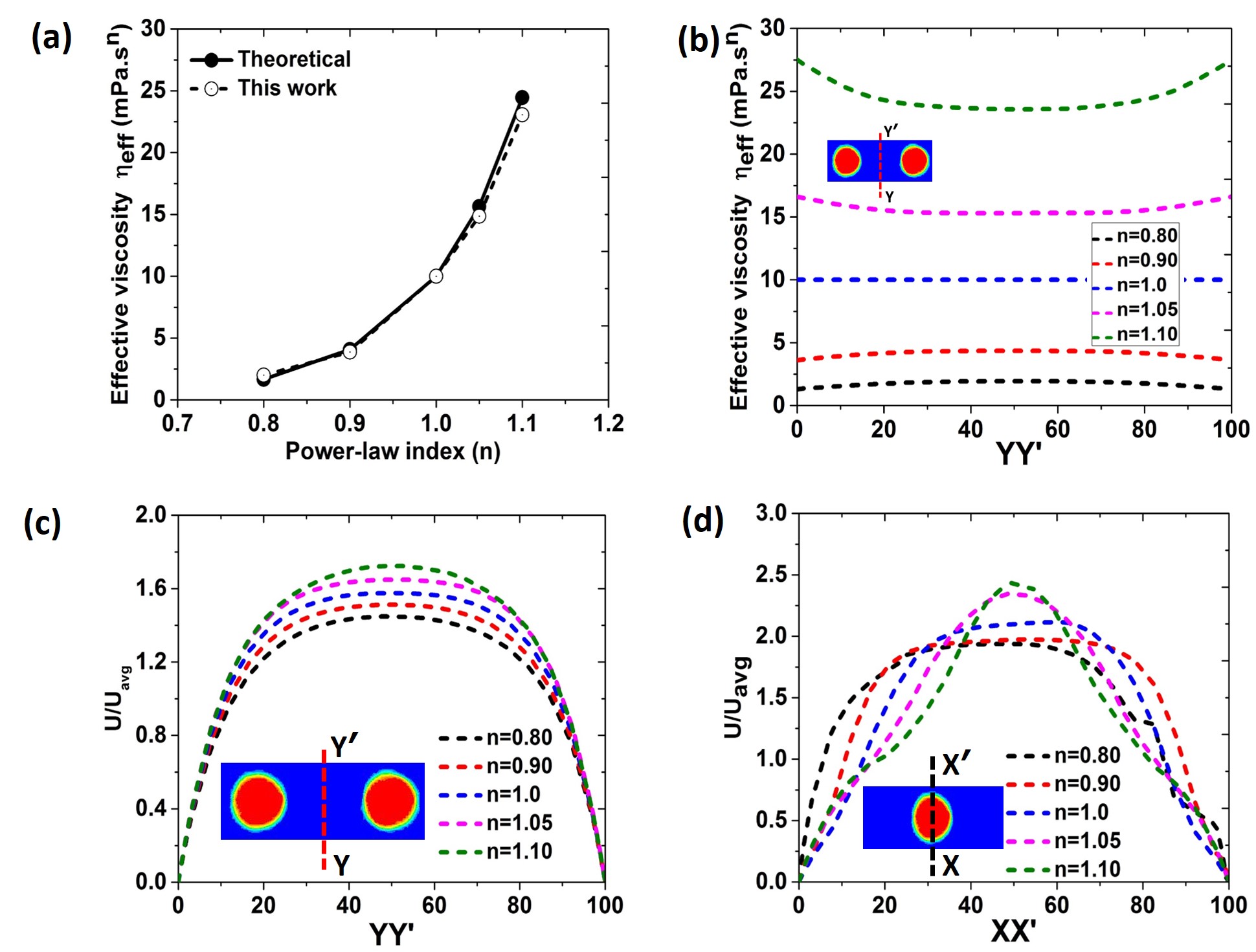}
	\caption{\label{fig:n41} (a) Effect of power\textendash law index on bulk liquid viscosity. Distribution profiles of (b) effective viscosity, and (c) velocity in middle of the liquid slug. (d) Velocity profiles in the middle of droplet at fixed $K= 0.01~Pa .s^n$, $\eta_w = 0.001~Pa.s$, $\sigma = 0.0365~N/m$, $Q_o = 0.408 ~\mu L/s$, and $Q_w = 0.14~\mu L/s$. }
\end{figure}  

To realize the effect of power-law index in terms of bulk viscosity and overall viscous force, effective viscosities of various liquids are estimated from the simulations and compared with the results derived from Eq.~\ref{eq:effective_eqn} \citep{rhodes-2008}. 

\begin{equation}
\label{eq:effective_eqn}
%\eta _{eff}  =K\left(\frac{8 U_{L} }{D}\right) ^{(n-1)}
\eta_{eff}=K\left(\frac{3n+1}{4n}\right)^{n}\left(\frac{8 U_{L} }{W_c}\right)^{n-1}
\end{equation}

where $K$, $U_L$, $W_c$, and $n$ are the consistency index, liquid inlet velocity, width of the channel, and power\textendash law index, respectively. Fig.~\ref{fig:n41}a shows enhanced effective viscosity with increasing power\textendash law index and the CFD calculations are in excellent agreement with the theoretical values. Effective viscosity distributions in the middle of the liquid slugs are also presented in Fig. \ref{fig:n41}b. In the middle of the microchannel, the effective viscosity increases for shear thinning liquids and decreases for shear thickening liquids which are the typical representation of non-Newtonian flow behavior. Velocity profiles in the middle of the slug and the droplet are also illustrated in Fig. \ref{fig:n41}c and Fig. \ref{fig:n41}d, respectively. In all cases, flatter velocity profiles are observed for shear thinning liquids, as expected. Due to enhanced film thickness near the wall and shape of the droplet, the velocity inside the droplet for shear thickening liquids is considerably higher than the shear thinning liquids.
	
	\begin{figure} [!ht]
		\centering
		\includegraphics[width=1\textwidth]{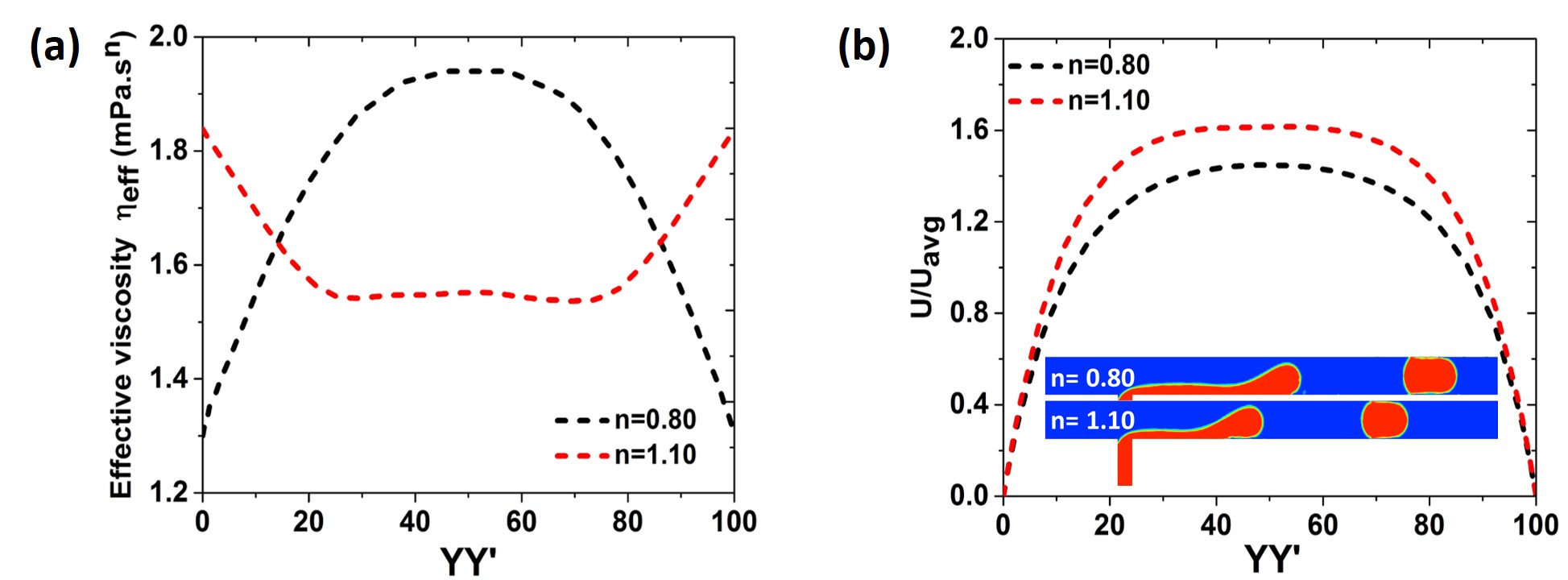}
		\caption{\label{fig:n51} Distribution profiles of (a) effective viscosity, and (b) velocity in middle of liquid slug in shear thinning ($n$=0.80 and $K= 0.01~Pa .s^n$) and shear thickening liquids ($n$=1.10 and $K= 0.00066~Pa .s^n$) at fixed effective viscosity $\eta_{eff}= 0.001669~Pa.s^n$,$\sigma = 0.0365~N/m$, $Q_o = 0.408 ~\mu L/s$, and $Q_w = 0.14~\mu L/s$.}
		
	\end{figure} 

To understand the rheological behavior of non-Newtonian liquids on the droplet formation, a set of simulations are performed by adjusting the value of $K$ which results in maintaining the identical effective viscosities for both shear thinning ($n$=0.80) and shear thickening ($n$=1.10) liquids. The effective viscosity distribution profiles in the middle of the liquid slug is shown in  Fig.~\ref{fig:n51}a. Similar to the earlier observation, typical shear dependent viscosity profiles are also apparent for both the liquids in this case. Dimensionless droplet length for shear thinning liquid ($L_D/W_C=1.860$) is found to be higher than shear thickening liquid ($L_D/W_C=1.575$) and the corresponding phase contours are shown in the inset of Fig. \ref{fig:n51}b. The velocity profiles illustrated in Fig. \ref{fig:n51}b also depict the combined influence of $n$ and $K$ even when the effective viscosities are same for both the liquids.

%For $n$\textless1, the pressure drop remains almost constant in the microchannel due to lower viscous stress. 
%
%However, for shear thickening fluids, the amount of dispersed phase pressure gradually increases during this process, which leads to a sudden droplet break \textendash up at the neck. 
%
%This sudden break leads to a small pressure jump as shown in Fig.\ref{fig:n2}d, . In Fig.\ref{fig:n2}d, each pulse indicates droplet position, . Typically the pressure drop exists at two sides of the droplet, due to change in radius of curvature. several  researcher reported . The present numerical analysis of pressure profiles showed reasonable agreement with literature data.  

\subsection{Effect of consistency index}
\begin{figure}[!ht]
	\centering
	\includegraphics[width=1\textwidth]{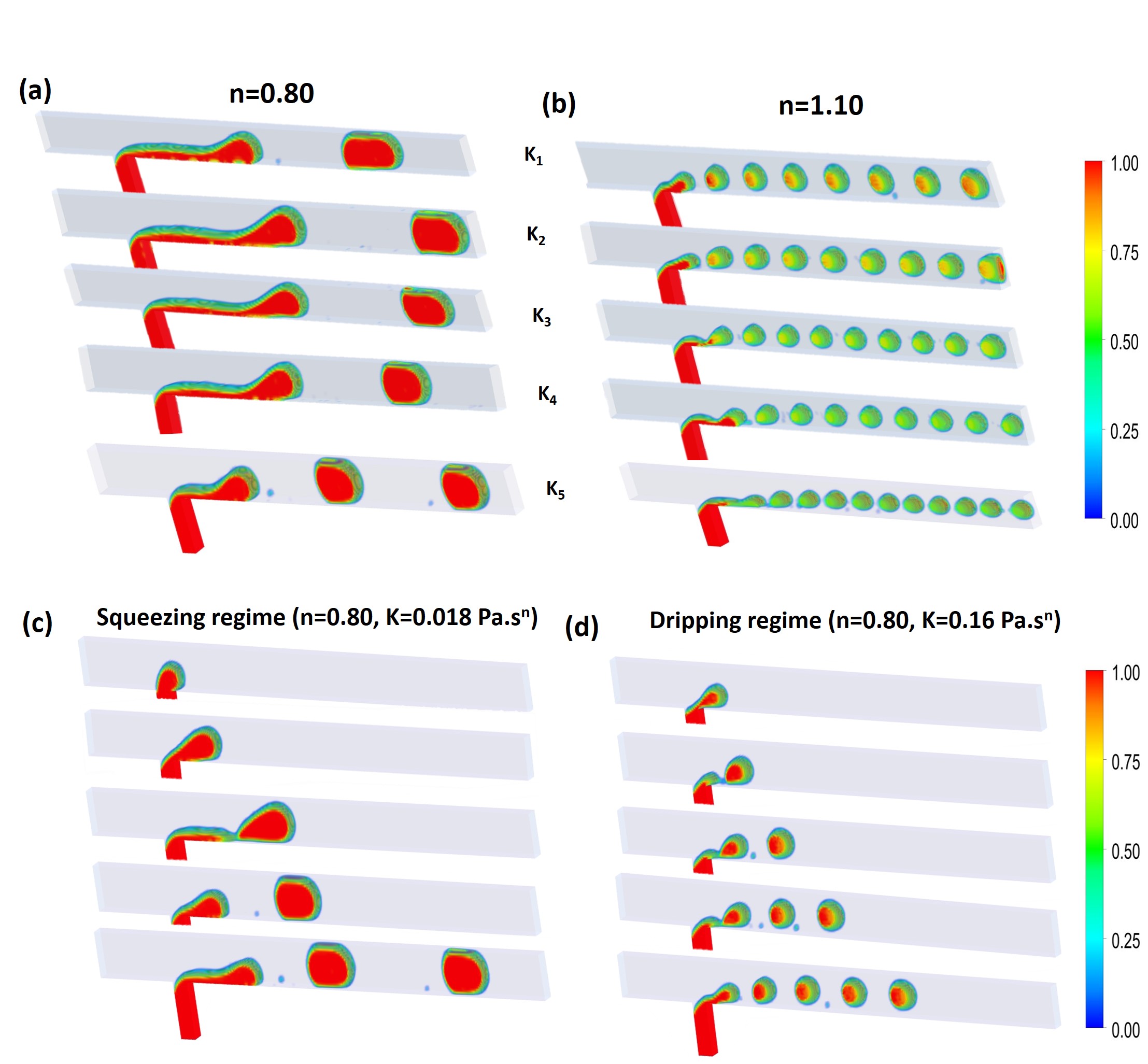}
	\caption{\label{fig:k1} Influence of $K$ during droplet formation in (a) shear thinning liquid ($n$=0.80), (b) shear thickening liquid ($n$=1.10) with $K_1= 0.008~Pa.s^n$, $K_2= 0.01~Pa.s^n$, $K_3= 0.012~Pa.s^n$, $K_4= 0.014~Pa.s^n$, $K_5= 0.018~Pa.s^n$. Droplet formation in a shear thinning liquid ($n$=0.80) during (c) squeezing mechanism with $K= 0.018~Pa.s^n$, and (d) dripping mechanism with $K= 0.16~Pa.s^n$ at a fixed operating condition of $Q_o = 0.408~\mu L/s$, $Q_w = 0.14~\mu L/s$, $\eta_w = 0.001~Pa.s$ and $\sigma = 0.0365~N/m$.}
\end{figure}

In this section, effect of consistency index ($K$) on droplet formation mechanism, length, and velocity has been methodically explored by altering $K$=0.008-0.018 $Pa.s^n$ of the power\textendash law liquids. Fig. \ref{fig:k1} shows the droplet formation mechanism in squeezing and dripping regimes. For the considered range of $K$, squeezing regime is experienced for $n$\textless1, and dripping mechanism is realized for $n$$\geq$1, as depicted in Fig. \ref{fig:k1}a and Fig. \ref{fig:k1}b, respectively. Typically, the forces acting on the dispersed phase at the merging junction are viscous, pressure difference, and interfacial tension forces. Viscous force is caused by the viscous stress acting on a liquid\textendash liquid interface and is proportional to the dispersed phase area along with the velocity gradient. For lower values of $K$, the dispersed phase is pulled downstream of the main channel before the droplets breakup due to dominant interfacial tension forces, as shown in Fig. \ref{fig:k1}a at $K=0.008~Pa.s^n$. For $n$\textless 1, interfacial and shear forces together provide squeezing action due to lower viscous forces. A long and thick layer of dispersed phase leads to the generation of elongated droplets for $n$=0.80 at $K$ = 0.008-0.014 $Pa.s^{n}$. However, further increase in consistency index value for $n$=0.80 and 0.90 results in flow regime transition from squeezing to dripping. Smaller droplets are observed for $n>1$ and flow regime shifts from squeezing to dripping due to the increase in effective viscosity, as discussed earlier. In a power law liquid, the droplet formation through squeezing as well as dripping mechanism can be realized by varying the consistency index value, as shown in Fig.~\ref{fig:k1}c and Fig.~\ref{fig:k1}d for $n=0.80$.
 
\begin{figure}[!ht]
	\centering
	\includegraphics[width=1\textwidth]{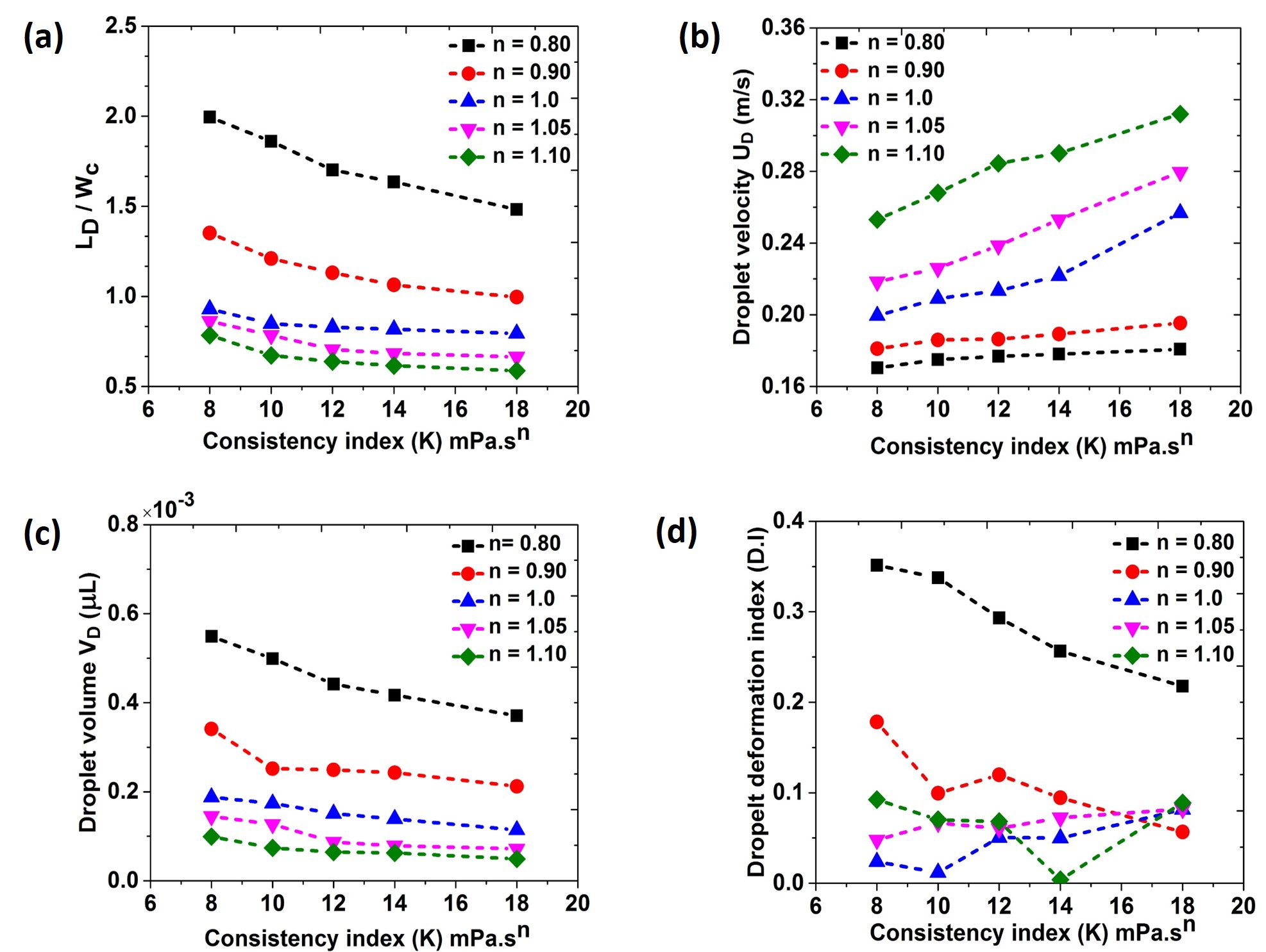}
	\caption{\label{fig:k2}   Effect of consistency index on droplet (a) length, (b) velocity (c) volume, and (d) deformation index at $Q_o = 0.408~\mu L/s$, $Q_w = 0.14~\mu L/s$, $\eta_w = 0.001~Pa.s$, and $\sigma = 0.0365~N/m$. }
\end{figure}

The droplet length is found to decrease with increasing $K$ as shown in Fig. \ref{fig:k2}a, due to increase in effective viscosity of the continuous phase liquid. For higher $K$ values, droplet length hardly changes for shear thickening liquids. From Fig. \ref{fig:k1}b, it can be noticed that under such scenario, droplet shape gradually changes toward nearly spherical (see for $K$ = 0.01 $Pa.s^n$) with reduced core diameter. Beyond this value of $K$, droplets are observed to fuse, attributing to extremely high viscous stress. Droplet velocity is also estimated to understand the effect of $K$ on droplet dynamics. 

\begin{figure}[!ht]
	\centering
	\includegraphics[width=0.80\textwidth]{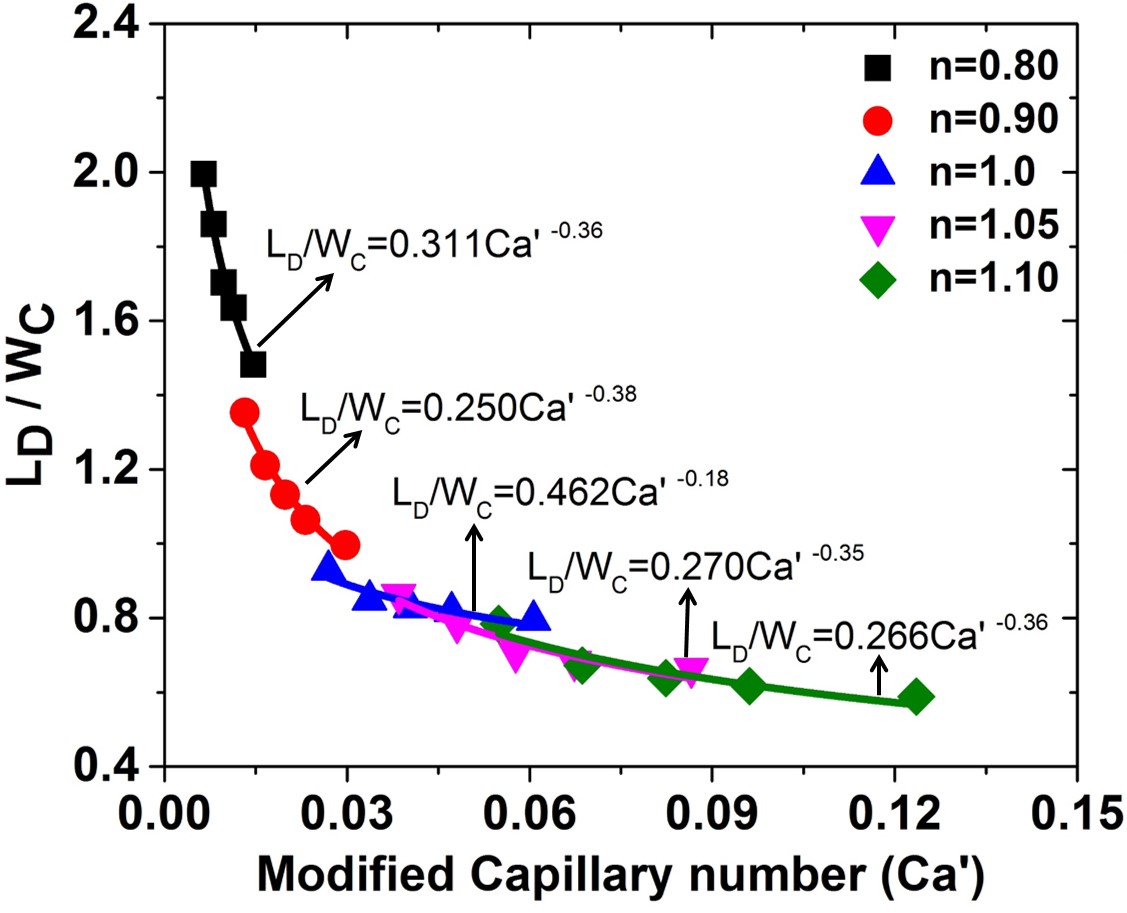}
	\caption{\label{fig:k213} The scaling relation for the non-dimensional droplet size with the modified Capillary number ($Ca^{'}$) for various power-law liquids at $Q_o = 0.408~\mu L/s$, $Q_w = 0.14~\mu L/s$, $\eta_w = 0.001~Pa.s$, and $\sigma = 0.0365~N/m$. }
\end{figure}

Fig. \ref{fig:k2}b depicts the gradual increase in droplet velocity with increasing $K$ which is more pronounced in shear thickening liquids. This is mainly due to increase in liquid film thickness around the droplets and reduction in droplet height ($H$ in Fig. \ref{fig:M}). Consequently, the droplet velocity marginally changes in shear thinning cases. However, a noticeable change in droplet velocity is observed for shear thickening and Newtonian liquids. Droplet volume is found to decrease with increase in $K$, as shown in Fig.~\ref{fig:k2}c. For higher $K$ values, droplet volume remains unchanged for shear thickening liquids. Deformation in shape is further quantified in Fig.~\ref{fig:k2}d which shows that, at lower consistency index ($K$=0.008-0.010 $Pa.s^n$), near spherical shaped droplets are observed for Newtonian liquids. For shear thinning liquid, plug shaped droplets are observed and the deformation decreases with increase in $K$. In the range of this study, it is observed that the droplet deformation index (D.I) hardly varies for Newtonian and shear thickening liquids. However, there is a dependence of D.I in cases of shear thinning liquids which becomes stronger with lower effective viscosity. This can be attributed to the lower shear force for droplet detachment from the neck of the junction in cases of shear thinning liquids. The non-dimensional size of the formed droplets ($L_D/W_c$) are scaled with the modified Capillary number  ($Ca^{'}$) as a power\textendash law relationship and are illustrated in Fig. \ref{fig:k213}. The proposed scaling laws for Newtonian and non\textendash Newtonian liquids in the range of $0.008 \leq K \leq 0.018$ $Pa.s^{n}$ are summarized in Table \ref{cr1} which have a maximum deviation of 5\%. It can be observed that the proposed scaling laws to predict droplet size are similar to those suggested for the droplet/bubble formation in Newtonian and non\textendash Newtonian liquids by several researchers \cite{xu-2008,fu2012bre,fu2016as,lu2014} in various microfluidic devices, but with different prefactors and exponents.

%These findings were quite surprising and suggest that further increase in consistency index for $n$\textgreater 1, lead to smaller and non\textendash uniform sized droplets. Furthermore, droplets fused into single droplet in the main channel due to a larger resistance in the continuous phase liquid. This study shows the critical  value of consistency index $K$ = 0.014 $Pa.s^n$ to generate droplets with mono\textendash dispersed size in continuous phase fluid ($n$ $\leq$ 1.10).

%In the  dripping mechanism, droplet length is smaller than channel dimensions with monodispersed droplets. 

\begin{table}[H]
\scriptsize 
%\footnotesize 
\centering
\caption{Scaling of the droplet size with consistency index for various power-law liquids}
\vspace{10px}
\label{cr1}
\begin{tabular}{lclll}
	\hline
\multicolumn{1}{c}{\begin{tabular}[c]{@{}c@{}}Power-law\\  index (n) \end{tabular}}  & \begin{tabular}[c]{@{}c@{}}Consistency index \\ $K$ ($Pa.s^{n}$) \end{tabular} & \multicolumn{1}{c}{\begin{tabular}[c]{@{}c@{}}Flow rates and \\ fluid properties\end{tabular}}   & \multicolumn{1}{c}{Scaling Laws} & \multicolumn{1}{c}{\begin{tabular}[c]{@{}c@{}}Modified Capillary \\ number ($Ca^{'}$)\end{tabular}} \\  \hline

	\multicolumn{1}{c}{0.80}                            & \multirow{5}{*}{0.008-0.018}                                          & \multirow{5}{*}{\begin{tabular}[c]{@{}l@{}} $Q_o = 0.408~\mu L/s$,\\ $Q_w = 0.14~\mu L/s$ and\\   $\sigma = 0.0365~N/m$\end{tabular}} & \multicolumn{1}{c}{$L_{D}/W_{C}=0.311 {{Ca^{'}}^{-0.36}}$}                          & \multicolumn{1}{c}{0.006-0.0146}                                       \\  
	
\multicolumn{1}{c}{0.90}                            &                                                                       &                                                                                        & \multicolumn{1}{c}{$L_{D}/W_{C}=0.250 {{Ca^{'}}^{-0.38}}$}                                  &  \multicolumn{1}{c}{0.0132-0.0297}                                               \\ 
\multicolumn{1}{c}{1.0}                             &                                                                       &                                                                                        &  \multicolumn{1}{c}{$L_{D}/W_{C}=0.462 {{Ca^{'}}^{-0.18}}$}                                 &  \multicolumn{1}{c}{0.0269-0.0606}                                               \\ 
\multicolumn{1}{c}{1.05}                            &                                                                       &                                                                                        &  \multicolumn{1}{c}{$L_{D}/W_{C}=0.270 {{Ca^{'}}^{-0.35}}$}                                 &  \multicolumn{1}{c}{0.03847-0.0865}                                               \\ 
\multicolumn{1}{c}{1.10}                            &                                                                       &                                                                                        &  \multicolumn{1}{c}{$L_{D}/W_{C}=0.266 {{Ca^{'}}^{-0.36}}$}                                 &  \multicolumn{1}{c}{0.0549-0.1235}                                               \\ \hline
\end{tabular}
\end{table}

\subsection{Effect of interfacial tension}
\begin{figure}[h]
	\centering
	\includegraphics[width=1\textwidth]{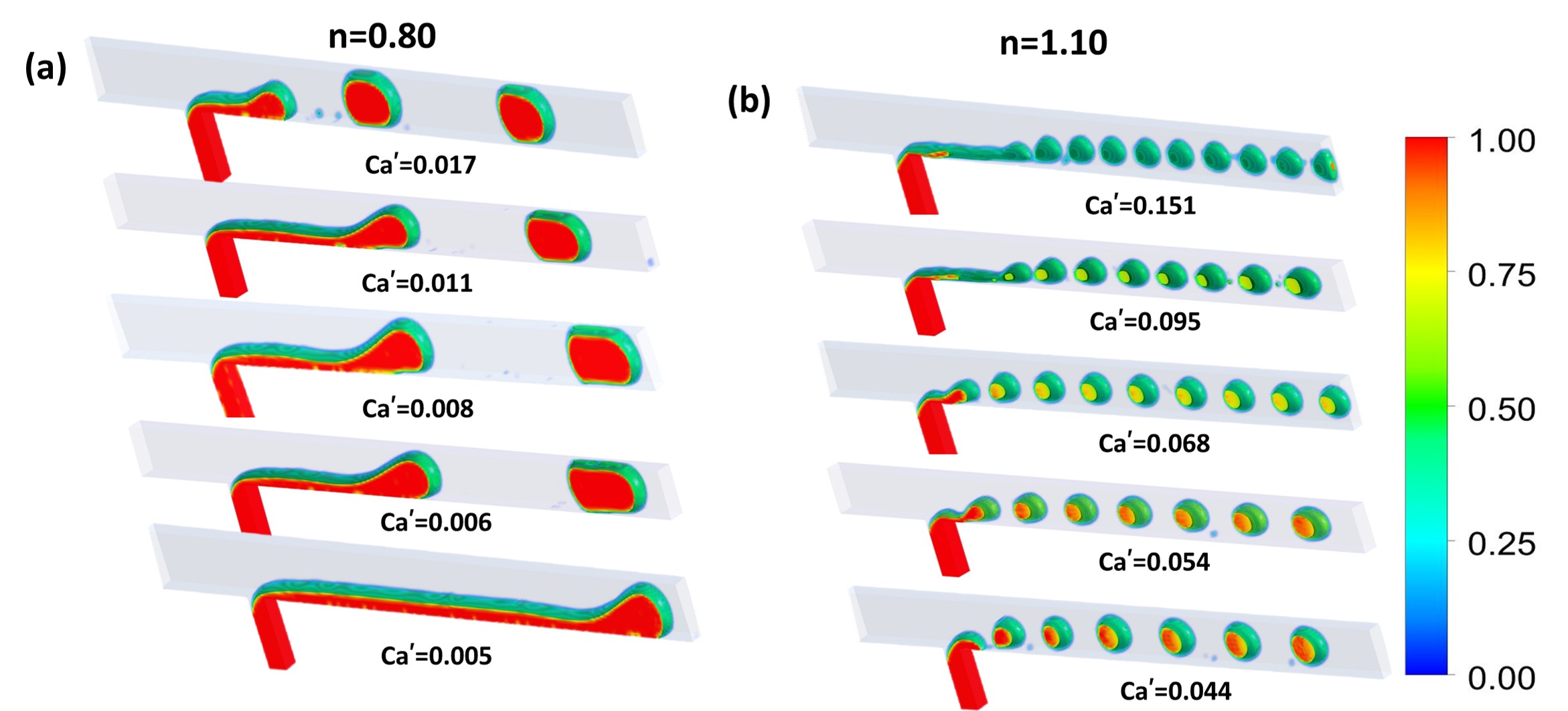}
	\caption{\label{fig:s1} Influence of interfacial tension on droplet formation for (a) shear thinning liquid ($n=0.80$) having $Ca^{'}=0.017$ ($\sigma = 0.0165~N/m$), $Ca^{'}=0.011$ ($\sigma = 0.0265~N/m$), $Ca^{'}=0.008$ ($\sigma = 0.0365~N/m$), $Ca^{'}=0.006$ ($\sigma = 0.0465~N/m$), and $Ca^{'}=0.005$ ($\sigma = 0.0565~N/m$), and (b) shear thickening liquid ($n=1.10$) having $Ca^{'}=0.151$ ($\sigma = 0.0165~N/m$),  $Ca^{'}=0.095$ ($\sigma = 0.0265~N/m$), $Ca^{'}=0.068$ ($\sigma = 0.0365~N/m$),  $Ca^{'}=0.054$ ($\sigma = 0.0465~N/m$) and $Ca^{'}=0.044$ ($\sigma = 0.0565~N/m$) at ${K}=0.01~Pa.s^{n}$, $\eta_w = 0.001~Pa.s$, $Q_o = 0.408~\mu L/s$, and $Q_w = 0.14~\mu L/s$.}
\end{figure}

To understand the effect of interfacial tension on the droplet length and velocity, a set of simulation is carried out keeping other properties constant. All the results are analyzed based on modified Capillary number $Ca^{'}$ ($=\frac{K U_L^{n} w_c^{1-n}}{ \sigma }$). It can be seen from Fig. \ref{fig:s1}a that for shear thinning liquid, flow regime shifts from dripping to jetting with increasing interfacial tension. At higher interfacial tension ($\sigma = 0.0565$ N/m and  ($Ca^{'}$=$0.005$), a thick layer of dispersed phase covers the channel length and droplet formation is not observed. Interestingly, for shear thickening ($n$=1.10) liquid, the flow regime shifts from small beads that are linked by a filament of the dispersed phase to dripping with increase in interfacial tension and is illustrated in Fig. \ref{fig:s1}b. Similar observation was reported by \citet{arratia2008p} in their experimental findings. This can be attributed to the fact that at lower interfacial tension, the growth of dispersed phase at the merging junction is hindered by the higher shear force, which in turn results into smaller droplet length for all cases. 

\begin{figure}[!ht]
	\centering
	\includegraphics[width=1\textwidth]{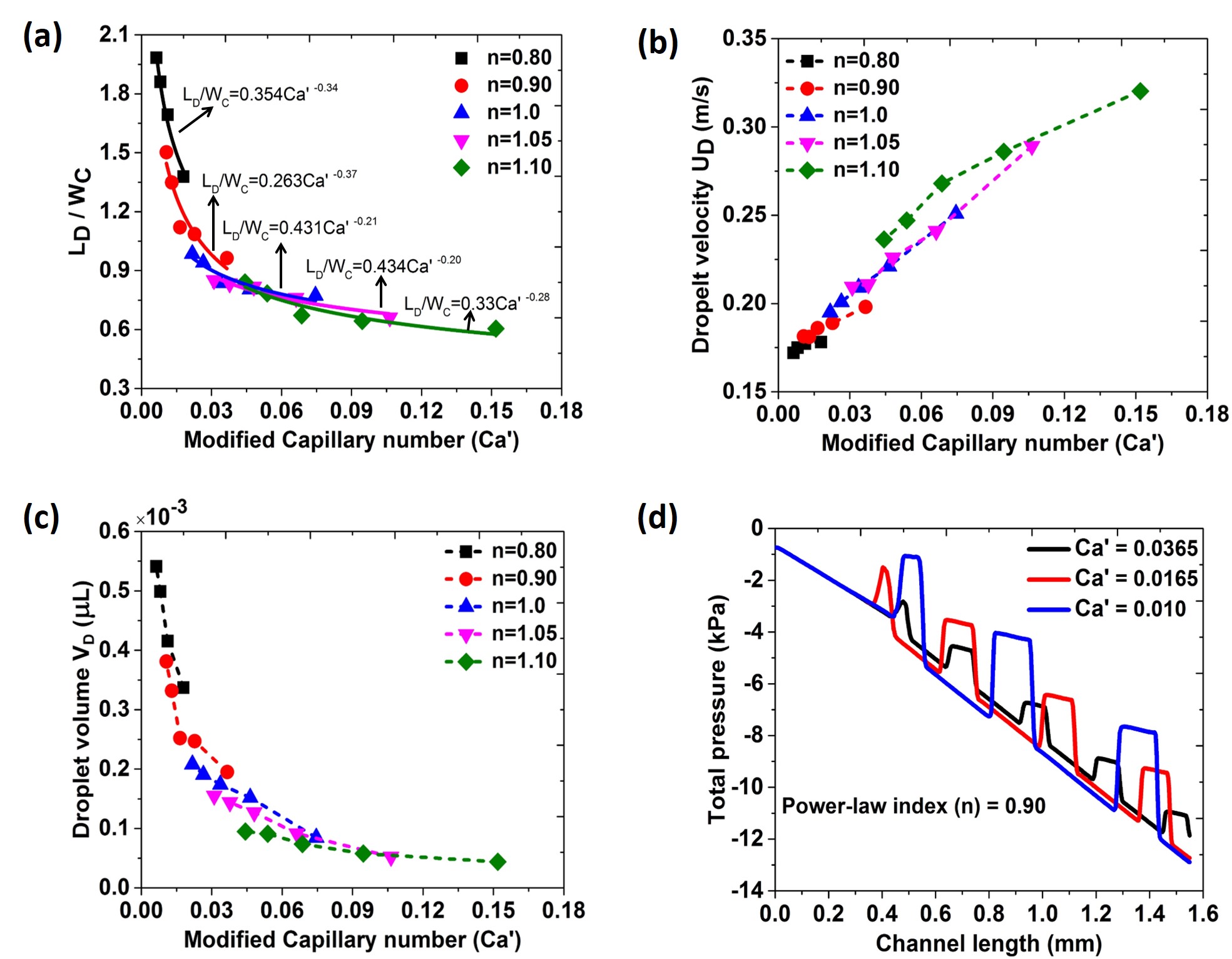}
	\caption{\label{fig:s2}Effect of interfacial tension on droplet (a) length, (b) velocity, (c) volume and (d) pressure profiles along the channel length at $Q_o = 0.408~\mu L/s$, $Q_w = 0.14~\mu L/s$, $K= 0.01~Pa.s^{n}$, and $\eta_w = 0.001~Pa.s$ in various power-law liquids. }
\end{figure}
  
Fig. \ref{fig:s2}a indicates the change in non-dimensional droplet length with varying $Ca^{'}$. It can be seen that with decreasing interfacial tension, the droplet length and height also decrease in all cases. Simple power law relations for various $n$ are proposed for the droplet length as a function of $Ca^{'}$ which are shown in Fig.~\ref{fig:s2}a. The applicability range of these relations are listed in Table \ref{cr2} which show a maximum error of 8\%. Consequently, considerable change in droplet velocity is observed, as shown in Fig. \ref{fig:s2}b. As droplet blocks the entire channel cross-section with increasing interfacial tension, its velocity decreases and this phenomena is pronounced in shear thickening cases unlike shear thinning liquid where the droplet height ($H$) remains constant. In line with earlier discussions, droplet volume is found to decrease with decreasing interfacial tension (Fig. \ref{fig:s2}c). Pressure drop across the droplet increases with increasing interfacial tension owing to change in curvature of the dispersed phase, as shown in Fig. \ref{fig:s2}d. However, the overall pressure drop across the channel length is constant.        
%
%The present study is qualitatively in-line with literature data \citep{wang-2011,liuu-2009}. These droplet velocity results are in good agreement with other studies as shown by \citet{van-2006}. 

%Pressure droplet also increases with increasing the interfacial tension shown in Fig.\ref{fig:s2}d, this is  due to change in curvature of dispersed phase as mentioned in earlier section. 

%From the literature, the breakup of systems with a high interfacial tension and a low viscosity of the jet phase falls into the inertial-dominated regime. In this regime, the inertial force dominates over the viscous force and is balanced by the interfacial force at the transition from dripping to jet 

\begin{table}[!ht]
	\scriptsize 
	%\footnotesize 
	\centering
	\caption{Scaling of the droplet size with interfacial tension for various power-law liquids}
	\vspace{10px}
	\label{cr2}
	\begin{tabular}{lclll}
		\hline
		\multicolumn{1}{c}{\begin{tabular}[c]{@{}c@{}}Power-law\\  index (n) \end{tabular}}  & \begin{tabular}[c]{@{}c@{}}Interfacial tension \\ $\sigma$ (N/m) \end{tabular} & \multicolumn{1}{c}{\begin{tabular}[c]{@{}c@{}}Flow rates and \\ fluid properties\end{tabular}}   & \multicolumn{1}{c}{Scaling Laws} & \multicolumn{1}{c}{\begin{tabular}[c]{@{}c@{}}Modified Capillary \\ number ($Ca^{'}$)\end{tabular}} \\  \hline

		\multicolumn{1}{c}{0.80}                            & \multirow{5}{*}{0.0165-0.0565}                                          & \multirow{5}{*}{\begin{tabular}[c]{@{}l@{}} $Q_o = 0.408~\mu L/s$,\\ $Q_w = 0.14~\mu L/s$ and\\   $K= 0.01~Pa.s^{n}$\end{tabular}} & \multicolumn{1}{c}{$L_{D}/W_{C}=0.354 {{Ca^{'}}^{-0.34}}$}                          & \multicolumn{1}{c}{0.0052-0.01796}                                       \\  
		
		\multicolumn{1}{c}{0.90}                            &                                                                       &                                                                                        & \multicolumn{1}{c}{$L_{D}/W_{C}=0.236 {{Ca^{'}}^{-0.37}}$}                                  &  \multicolumn{1}{c}{0.0106-0.0365}                                               \\ 
		\multicolumn{1}{c}{1.0}                             &                                                                       &                                                                                        &  \multicolumn{1}{c}{$L_{D}/W_{C}=0.431 {{Ca^{'}}^{-0.21}}$}                                 &  \multicolumn{1}{c}{0.0217-0.0745}                                               \\ 
		\multicolumn{1}{c}{1.05}                            &                                                                       &                                                                                        &  \multicolumn{1}{c}{$L_{D}/W_{C}=0.434 {{Ca^{'}}^{-0.20}}$}                                 &  \multicolumn{1}{c}{0.0310-0.1063}                                               \\ 
		\multicolumn{1}{c}{1.10}                            &                                                                       &                                                                                        &  \multicolumn{1}{c}{$L_{D}/W_{C}=0.33 {{Ca^{'}}^{-0.28}}$}                                 &  \multicolumn{1}{c}{0.0443-0.1518}                                               \\ \hline
	\end{tabular}
\end{table}

\subsection{Effect of flow rate}     
\subsubsection{Continuous phase}
\begin{figure}[H]
	\centering
	\includegraphics[width=1\textwidth]{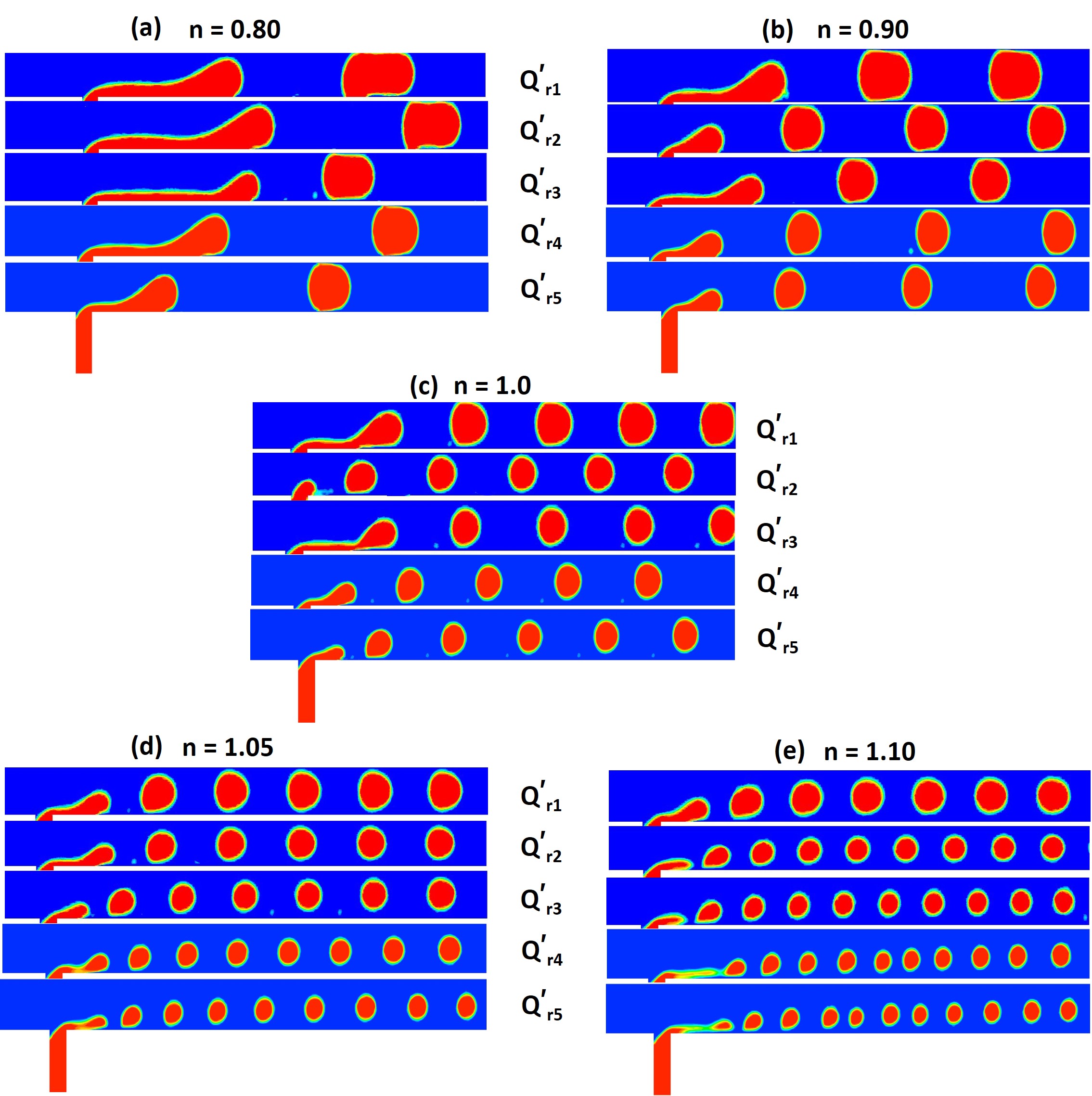}
	\caption{\label{fig:v1} Effect of continuous phase to dispersed phase flow rate ratio ($Q^{'}_r$) on droplet formation in (a) n = 0.80, (b) n = 0.90, (c) n = 1.0, (d) n = 1.10, and (e) $n$=1.10 for different flow rate ratios of $Q^{'}_{r1} = 0.212 $ $\mu L/s$, $Q^{'}_{r2} = 2.915 $ $\mu L/s$, $Q^{'}_{r3} = 3.536 $ $\mu L/s$, $Q^{'}_{r4} = 4.243 $ $\mu L/s$ and $Q^{'}_{r5} = 4.95 $ $\mu L/s$ at a fixed  operating condition of ${ K}$ = 0.01 $Pa.s^{n}$, $\eta_w = 0.001$ Pa.s, $\sigma = 0.0365$ N/m and $ Q_{w} = 0.14 $ $\mu L/s$.}
\end{figure}

At a fixed operating condition of  ${K}$  = 0.01 $Pa.s^{n}$, $\sigma = 0.0365$ N/m, and $Q_{w} = 0.14 $ $\mu L/s$, the effect of continuous phase flow rate on droplet formation mechanism, length and deformation index is investigated. For a range of continuous flow rates from $Q_o = 0.297 $ $\mu L/s$ to $Q_o = 0.693 $ $\mu L/s$, only squeezing regime is observed for shear thinning liquid (Fig. \ref{fig:v1}a and Fig. \ref{fig:v1}b). However, transition from squeezing to dripping regime is observed for Newtonian liquid, as described in Fig. \ref{fig:v1}c. From Fig.~\ref{fig:v1}d and Fig.~\ref{fig:v1}e, it is apparent that in shear thickening liquid, dripping mechanism is dominant with increase in continuous phase flow rate. At lower flow rates of shear thinning continuous phase, the dispersed phase pushes easily into the main channel due to lower resistance in continuous phase and completely blocks the cross section leading to squeezing flow regime. The dispersed phase thread gradually decreases with increasing shear stress and the detachment occurs on attainment of critical thickness at the neck. 

%Typically dispersed phase gradually changes to thinning at neck due to the shear stress of continuous phase on the dispersed phase. Consequently, a gradual change in droplet length is observed in the dripping regime for shear thickening fluids.  

%At fixed fluid properties on increasing continuous flow rate from $Q_o = 0.297 $ $\mu L/s$ to $Q_o = 0.495 $ $\mu L/s$  and Newtonian liquids, this is  due to a greater influence of pressure build up more than shear and interfacial forces during droplet formation for shear thinning and Newtonian liquids as depicted in Fig.\ref{fig:v1}a,b and c. 

%It is interesting to note that, flow transition is observed  from squeezing to the dripping mechanism with increasing the continuous phase flow rate for Newtonian to shear thickening liquids. Similar flow transition observations reported in the open literature by several authors \citep{de-2008,glawde-2012}. 

%Because, at higher $n$ values, viscous stress increases in continuous phase liquid, thus leading to a smaller length of droplets pinch off at the neck. 

\begin{figure}[!ht]
	\centering
	\includegraphics[width=1\textwidth]{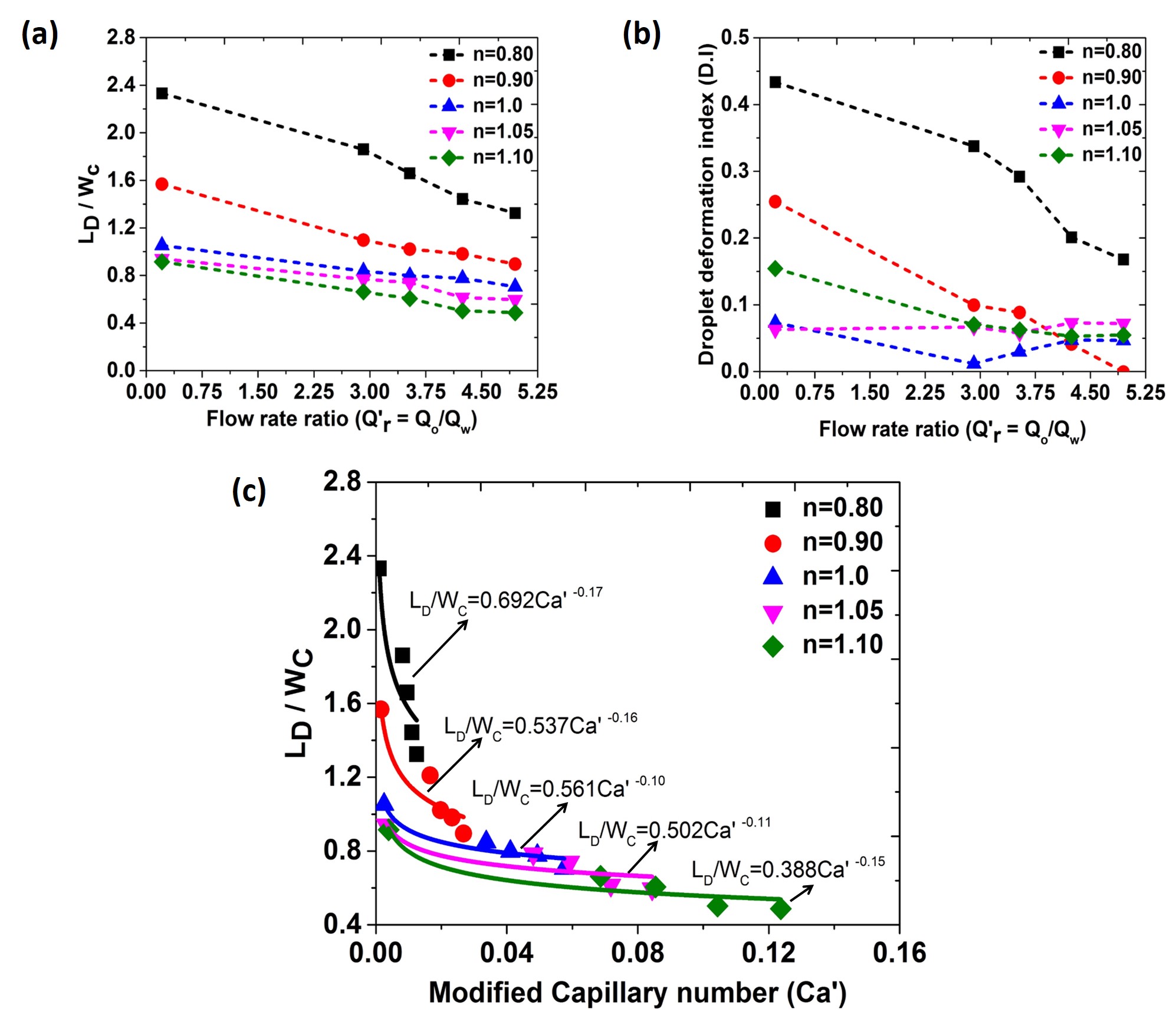}
	\caption{\label{fig:v2} Effect of continuous phase flow rate on (a) non-dimensional droplet length, and (b) droplet deformation index at a fixed operating conditions of ${ K}$  = 0.01 $Pa.s^{n}$, $\sigma = 0.0365$ N/m, $\eta_w = 0.001$ Pa.s and  $ Q_{w} = 0.14 $ $\mu L/s$. (c) The scaling of non-dimensional droplet length with $Ca^{'}$ for different continuous phase flow rates in power-law liquids.}
\end{figure}

However, with increasing $n$ and continuous phase flow rate, increased viscous and inertia forces result in rapid detachment of the droplet at the neck, as evident from some cases of Newtonian and all cases of shear thickening liquid. For shear thickening liquid, greater resistance is imparted by the continuous phase liquid causing the dispersed phase to grow slowly into the main channel until it balances all forces at the junction and eventually leads to reduction in droplet length (Fig. \ref{fig:v2}a). Similar to the previous observation, near spherical and plug shaped droplets are observed in dripping and squeezing regimes, respectively (Fig.~\ref{fig:v2}b). Fig.~\ref{fig:v2}c shows power law relationship between the non-dimensional droplet size and $Ca^{'}$ resulting from the variation of continuous phase flow rate. The proposed relations are listed in Table \ref{cr3} for 0.297 $\leq$ $Q_{o}$ $\leq$ 0.693 $\mu$L/s and predicts with a maximum deviation of 3\%.

\begin{table}[!ht]
	\scriptsize 
	%\footnotesize 
	\centering
	\caption{Scaling of the droplet size with continuous phase flow rate for various power-law liquids}
	\vspace{10px}
	\label{cr3}
	\begin{tabular}{lclll}
		\hline
		\multicolumn{1}{c}{\begin{tabular}[c]{@{}c@{}}Power-law\\  index (n) \end{tabular}}  & \begin{tabular}[c]{@{}c@{}}Flow rates \end{tabular} & \multicolumn{1}{c}{\begin{tabular}[c]{@{}c@{}} Fluid properties\end{tabular}}   & \multicolumn{1}{c}{Scaling Laws} & \multicolumn{1}{c}{\begin{tabular}[c]{@{}c@{}}Modified Capillary \\ number ($Ca^{'}$)\end{tabular}} \\  \hline

		\multicolumn{1}{c}{0.80}                            &  \multirow{5}{*}{\begin{tabular}[c]{@{}c@{}}$Q_o =0.297-0.693~\mu L/s$ \\ $Q_w =0.14~\mu L/s$ \end{tabular}}                                      & \multirow{5}{*}{\begin{tabular}[c]{@{}l@{}} $\sigma = 0.0365$ N/m \\   $K= 0.01~Pa.s^{n}$\end{tabular}} & \multicolumn{1}{c}{$L_{D}/W_{C}=0.692 {{Ca^{'}}^{-0.17}}$}                          & \multicolumn{1}{c}{0.001-0.0124}                                       \\  
		
		\multicolumn{1}{c}{0.90}                            &                                                                       &                                                                                        & \multicolumn{1}{c}{$L_{D}/W_{C}=0.537 {{Ca^{'}}^{-0.16}}$}                                  &  \multicolumn{1}{c}{0.0015-0.0267}                                               \\ 
		\multicolumn{1}{c}{1.0}                             &                                                                       &                                                                                        &  \multicolumn{1}{c}{$L_{D}/W_{C}=0.561 {{Ca^{'}}^{-0.10}}$}                                 &  \multicolumn{1}{c}{0.0024-0.0575}                                               \\ 
		\multicolumn{1}{c}{1.05}                            &                                                                       &                                                                                        &  \multicolumn{1}{c}{$L_{D}/W_{C}=0.502 {{Ca^{'}}^{-0.11}}$}                                 &  \multicolumn{1}{c}{0.003-0.0843}                                               \\ 
		\multicolumn{1}{c}{1.10}                            &                                                                       &                                                                                        &  \multicolumn{1}{c}{$L_{D}/W_{C}=0.388 {{Ca^{'}}^{-0.15}}$}                                 &  \multicolumn{1}{c}{0.0038-0.1236}                                               \\ \hline
	\end{tabular}
\end{table}

\subsubsection{Dispersed phase}
%This section deals the  effect of dispersed phase flow rate on droplet length and deformation index for Newtonian and non\textendash Newtonian liquids at fixed operating conditions of  ${ K}$  = 0.01 $Pa.s^{n}$,$\gamma = 0.0365$ N/m and  $ Q_{o} = 0.408 $ $\mu L/s$.
Keeping the continuous phase flow rate constant at $Q_{o} = 0.408 $ $\mu L/s$, dispersed phase flow rate ($Q_w$) is varied from 0.14  $\mu L/s$ \textendash \ 0.539  $\mu L/s$ to understand its effect on droplet formation characteristics.
%
%, droplet length increases for Newtonian and non \textendash Newtonian liquids. 
%
%
%Influence of dispersed phase velocity is quantified in terms of dimensionless droplet length and droplet deformation index based on dispersed phase Reynolds number ($Re_w= \frac{D_w U_w\rho_w}{\mu_w}$).     

\begin{figure}[!ht]
	\centering
	\includegraphics[width=1\textwidth]{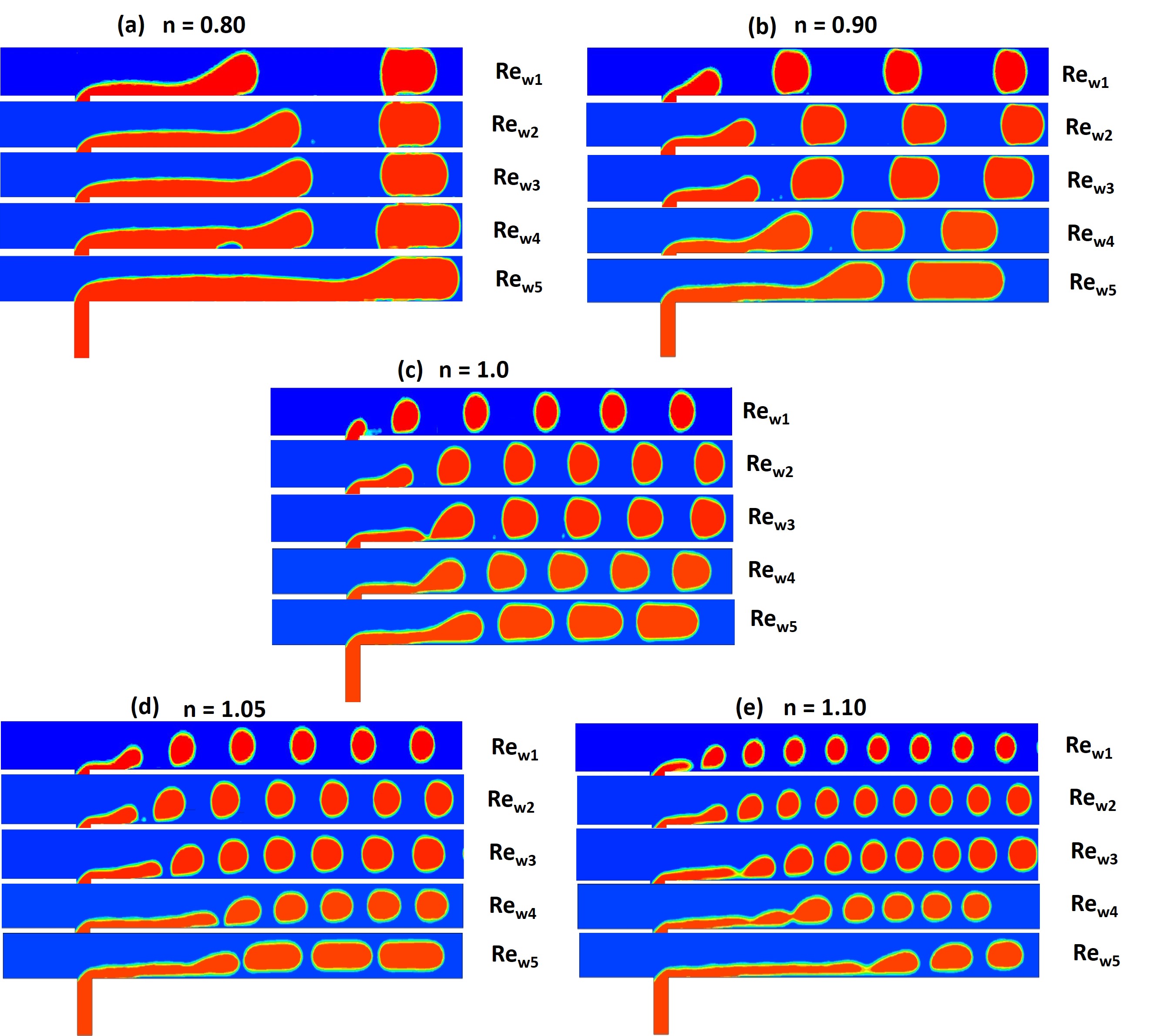}
	\caption{\label{fig:vd1}  Effect of dispersed phase flow rate ($Q_w$) on droplet formation in (a) n = 0.80 (b) n = 0.80 (c) n = 1.05 and (d) n = 1.10 for different dispersed phase Reynolds numbers: $Re_{w1}$=4.24, $Re_{w2}$=6.30, $Re_{w3}$=8.45, $Re_{w4}$=10.3 and $Re_{w5}$=16.36 at ${ K}$  = 0.01 $Pa.s^{n}$, $\eta_w = 0.001$ Pa.s, $\sigma = 0.0365$ N/m and $ Q_{o} = 0.408 $ $\mu L/s$. }
\end{figure}

\begin{figure}[!ht]
	\centering
	\includegraphics[width=1\textwidth]{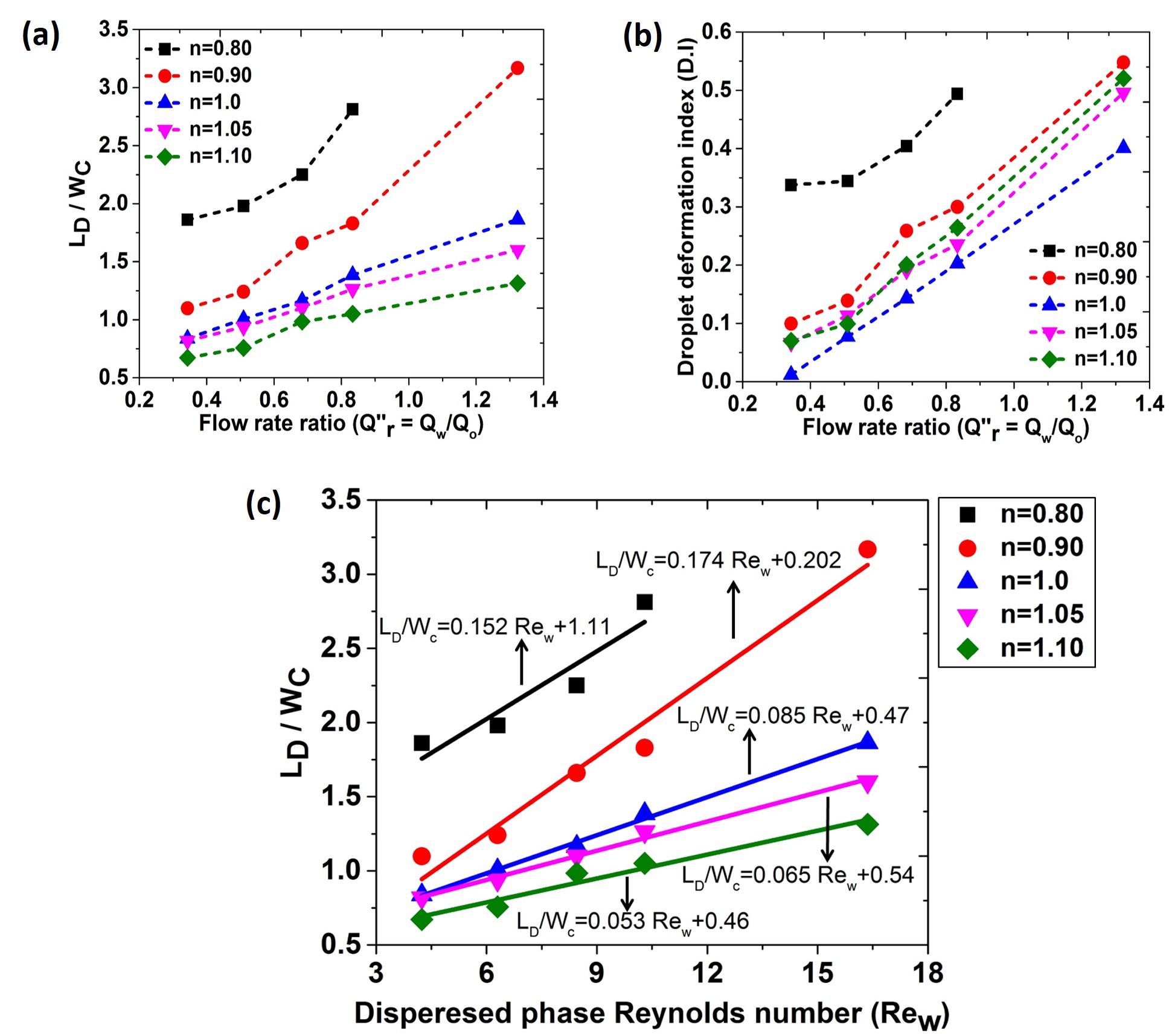}
	\caption{\label{fig:vd2} Effect of dispersed phase Reynolds number on droplet (a) length and (b) droplet formation index ($D.I$) at ${ K}$  = 0.01 $Pa.s^{n}$, $\eta_w = 0.001$ Pa.s, $\sigma = 0.0365$ N/m and  $ Q_{o} = 0.408 $ $\mu L/s$. (c) The scaling of non-dimensional droplet length with $Re_w$ for different dispersed phase flow rates in power-law liquids.  }
\end{figure} 

The results are explained in terms of dispersed phase Reynolds number ($Re_w= \frac{D_w U_w\rho_w}{\eta_w}$). From Fig. \ref{fig:vd1}, it can be observed that for shear thinning cases, squeezing regime prevails with increasing $Q_w$ (Figs. \ref{fig:vd1}a and \ref{fig:vd1}b). However, for Newtonian liquid, the regime shifts from dripping to squeezing with increasing $Q_w$ due to increased amount of dispersed phase being pushed into the channel at a fixed operating condition (Fig. \ref{fig:vd1}c). In case of shear thickening liquid, dripping regime is observed which moves toward jetting regime with increased $Q_w$ (Figs. \ref{fig:vd1}d and \ref{fig:vd1}e) due to the combined effect of higher inertia force (from dispersed phase) and viscous stress (in continuous phase).  Consequently, the droplet length increases significantly in shear thinning liquid with increasing dispersed phase velocity (Fig. \ref{fig:vd2}a) as the inertia force is significant enough to resist the opposing continuous phase shear stress. However, in Newtonian and shear thickening liquid, that inertia effect is suppressed by continuous phase shear and viscous stresses resulting in small change in droplet length. Due to increase in droplet volume with increasing $Q_w$, the droplet shape also changes from near spherical to plug type, as indicated by droplet deformation index in Fig. \ref{fig:vd2}b. For various $n$, linear scaling relationships are proposed between non-dimensional droplet size and the dispersed phase Reynolds number ($Re_{w}$) as shown in Fig.\ref{fig:vd2}c. The range of applicability is provided in Table \ref{cr4}.

%From  Fig.\ref{fig:vd2}a the effect of  $Re_w$ reveal that, droplet size increases with dispersed phase velocity. This due to increase in dispersed phase velocity( $Re_w$=4.28\textendash 8.45) higher  inertia force creates in the dispersed phase during droplet formation process. Dispersed phase . Continuous phase   Shear force helps to move two fluids interface towards T\textendash junction corner where droplet pinch-off occurs.

%\noindent In the case of shear thinning liquids, droplet length is higher than Newtonian and shear thickening liquids due effective viscosity. Droplet length found to be almost same for Newtonian and shear thickening liquid (n=1.05). Substantial change in droplet length is  found for higher shear thickening liquid (n=1.10) due to higher in effective viscosity. The droplet deformation index  also increases with increasing the dispersed phase velocity as shown in \ref{fig:vd2}b. This result reveals that, at higher dispersed phase velocity plug shape droplets are formed. Fig.\ref{fig:vd2}b indicates that droplet deformation index slightly changes between $Re_w$= 4.24\textendash 6.3. Deformation index  gradually increases beyond $Re_{w2}$=6.3 its means that, droplets shape deviating form spherical to deformed state . 

\begin{table}[!ht]
	\scriptsize 
	%\footnotesize 
	\centering
	\caption{Scaling of the droplet size with dispersed phase flow rate for various power-law liquids}
	\vspace{10px}
	\label{cr4}
	\begin{tabular}{lclll}
		\hline
		\multicolumn{1}{c}{\begin{tabular}[c]{@{}c@{}}Power-law\\  index (n) \end{tabular}}  & \begin{tabular}[c]{@{}c@{}}Flow rates \end{tabular} & \multicolumn{1}{c}{\begin{tabular}[c]{@{}c@{}} Fluid properties\end{tabular}}   & \multicolumn{1}{c}{Scaling Laws} & \multicolumn{1}{c}{\begin{tabular}[c]{@{}c@{}} Reynolds \\ number ($Re_w$)\end{tabular}} \\  \hline

		\multicolumn{1}{c}{0.80}                            &  \multirow{5}{*}{\begin{tabular}[c]{@{}c@{}}$Q_w=0.14-0.539~\mu L/s$ \\ $Q_o =0.408~\mu L/s$ \end{tabular}}                          & \multirow{5}{*}{\begin{tabular}[c]{@{}l@{}} $\sigma = 0.0365$ N/m \\   $K= 0.01~Pa.s^{n}$\end{tabular}} & \multicolumn{1}{c}{$L_{D}/W_{C}=0.152 Re_{w}+1.11$}                          & \multirow{5}{*}{4.24-16.36}                                       \\  
		
		\multicolumn{1}{c}{0.90}                            &                             &             & \multicolumn{1}{c}{$L_{D}/W_{C}=0.174 Re_{w}+0.202$}                                  &                                               \\ 
		\multicolumn{1}{c}{1.0}                             &                            &             &  \multicolumn{1}{c}{$L_{D}/W_{C}=0.085 Re_{w}+0.47$}                                 &                                                 \\ 
		\multicolumn{1}{c}{1.05}                            &                            &             &  \multicolumn{1}{c}{$L_{D}/W_{C}=0.065 Re_{w}+0.54$}                                &                                                \\ 
		\multicolumn{1}{c}{1.10}                            &                          &           &  \multicolumn{1}{c}{$L_{D}/W_{C}=0.053 Re_{w}+0.46$}                                &                                                 \\ \hline
	\end{tabular}
\end{table}

\section{Conclusion}
A comprehensive computational study of Newtonian droplet formation in non-Newtonian power-law liquids is carried out in a T-junction microchannel using VOF method. New insights are obtained regarding the droplet formation process in non-Newtonian liquids. With increasing power-law and consistency index, droplet length is found to decrease as effective viscosity increases. The droplet length also decreases with increasing continuous phase flow rate. Squeezing, dripping, and jetting regimes are found to strongly depend on the flow rate ratio, interfacial tension, and rheological properties. A parameter defined as droplet deformation index to identify the shape of droplets, shows that near spherical droplets are typically formed in all cases of dripping and jetting regimes. However, plug shaped droplets are obtained in squeezing regime. Like Newtonian medium, it is observed that interfacial tension has significant influence on the droplet formation pattern and size in non-Newtonian liquids. With increasing interfacial tension, droplet size increases in all cases, however, the regime shifts from filament linked small beads to dripping in shear thickening case. 

The development of microfluidic methods to generate and manipulate monodisperse droplets have led to an increasing number of potentially interesting applications. Although the computational study can not completely diminish the necessity of exhaustive and expensive, at times, experimental investigation; a well-validated CFD model can certainly complement with various aspects of physical phenomena that are not attainable by experiments. Our findings are expected to provide better understanding and experimental guidelines in terms of controlling parameters on the formation of desired Newtonian droplet of different shape and size in a non-Newtonian medium. 

\section*{Acknowledgment}
This work is supported by Sponsored Research \& Industrial Consultancy (SRIC), IIT Kharagpur under the scheme for ISIRD (Code: FCF). 

\section*{Nomenclature}

\begin{longtable}{l p{12cm}}
	
	$Re_w$    & Reynolds number ($= \frac{D_w U_w\rho_w}{\mu_w}$)  \\
	$Ca^{'}$    & Modified Capillary number ($=\frac{K U_L^{n} W_c^{1-n}}{ \gamma }$)  \\
	$U$	    & velocity (m/s) \\
	$v$ 	& average velocity in a cell (m/s)\\
	$W$ 	& width of channel\\ 
	$H$		& height of the droplet\\
	$D.I$	& droplet deformation index\\
	$t$ 	& time (s)\\
	$P$		& pressure (Pa)\\
	$F_{SF}$ & volumetric surface tension force (N/m$^3$)\\
	 $K$  & consistency coefficient ($Pa.s^{n}$)\\
	 $Q$	& volumetric flow rate ($\mu L/s$) \\
	$\overline{\overline D}$ & rate\textendash of\textendash deformation tensor \\
    $Q^{'}_{r}$ & flow rate ratio ($Q_o/Q_w$)\\
    
	\textit{Greek symbol}\\
	
	$\alpha$  & volume fraction\\
	$\bigtriangleup t$ & time step (s) \\
	$\bigtriangleup x$ & cell size (m) \\      
	$\eta$ &  viscosity (Pa.s)\\
	$\theta$ & contact angle (\textdegree) \\	
	%$\mu _{ \circ }$ & apparent viscosity\\
	$\rho$  & density ($kg/m^{3}$)\\
	$\sigma$  & surface tension (N/m)\\  
	$\dot{ \gamma}$ & shear rate \\    
	$\kappa_{n}$ 	& radius of curvature\\
	$\eta _{ eff }$ 	& effective viscosity ($Pa.s$)\\
	$\overline{\overline \tau}$ & stress tensor \\
	
	\textit{Subscripts}\\

	$w$ & water\\
    $c$ & continuous phase\\
    $d$ & dispersed phase\\
    $D$ & droplet\\ 
	$o$ & oil\\     
	
\end{longtable}

\section*{References}
\bibliography{mybibfile}
\end{document}